\documentclass{article}
\usepackage[utf8]{inputenc}
\usepackage{graphicx}
\usepackage{authblk}
\usepackage{amsmath}
\usepackage{amssymb}
\usepackage[margin=1in]{geometry}

\usepackage[backend=biber,style=numeric,sortcites,natbib=true,sorting=none]{biblatex} 
\usepackage{tikz}
\newcommand*\circled[1]{\tikz[baseline=(char.base)]{
    \node[shape=circle, draw, inner sep=1pt, 
        minimum height=12pt] (char) {#1};}}
        
\title{Molten-globule like transition state of protein barnase measured with calorimetric force spectroscopy}

\author[1]{Marc Rico-Pasto}
\author[2]{Annamaria Zaltron}
\author[3]{Sebastian J. Davis}
\author[4]{Silvia Frutos}
\author[1]{Felix Ritort}

\affil[1]{Small Biosystems Lab, Condensed Matter Physics Department, University of Barcelona, C/Marti i Franques 1, 08028 Barcelona, Spain.}

\affil[2]{Physics and Astronomy Department, University of Padova, Via Marzolo 8, 35131 Padova, Italy.}

\affil[3]{Laboratory of Nanoscale Biology, Institute of Bioengineering, School of Engineering, EPFL, 1015 Lausanne, Switzerland.}

\affil[4]{ProteoDesign, Barcelona Science Park, Baldiri Reixac 10-12, 08028 Barcelona, Spain.}

\begin{document}

\maketitle

\begin{abstract}
Understanding how proteins fold into their native structure is a fundamental problem in biophysics, crucial for protein design. It has been hypothesized that the formation of a molten-globule intermediate precedes folding to the native conformation of globular proteins; however, its thermodynamic properties are poorly known. We perform single-molecule pulling experiments of protein barnase in the range of 7- 37ºC using a temperature-jump optical trap. We derive the folding free energy, entropy and enthalpy, and the heat capacity change ($\Delta C_p$ = 1050$\pm$50 cal/mol$\cdot$K) at low ionic strength conditions. From the measured unfolding and folding kinetic rates, we also determine the thermodynamic properties of the transition state, finding a significant change in $\Delta C_p$ ($\sim$90\%) between the unfolded and the transition state. In contrast, the major change in enthalpy ($\sim$80\%) occurs between the transition and native state. These results highlight a transition state of high energy and low configurational entropy structurally similar to the native state, in agreement with the molten-globule hypothesis.
\end{abstract}

Protein folding stands as one of the major open questions in biophysics. In the 50-60's decade, Anfinsen introduced the thermodynamic hypothesis claiming that proteins spontaneously fold to a free energy minimum under appropriate conditions \cite{anfinsen1961,anfinsen1973}. In 1969, Levinthal noticed a polypeptide chain could not fold into the native state by random search in configurational space \cite{levinthal1968}. Protein folding is akin to finding a needle in a haystack and must be driven by molecular forces \cite{dill1990}. In an effort to solve the paradox, Ptitsyn proposed the molten-globule hypothesis (MGH) where folding is similar to solid formation from a gas: a molten-globule state must precede protein folding, similarly to the metastable liquid phase preceding solid formation during gas deposition \cite{ptitsyn1995}. The dry molten-globule is a necessary intermediate \cite{Hua2008, Jha2014} to form the native state (hereafter denoted as N) that is structurally similar to it but with most native bonds not yet formed. For years scientists have searched for folding intermediates, the most natural solution to Levinthal's paradox. While these have been identified in large proteins, many small globular proteins fold in a two-states manner, raising the question whether such  molten-globule intermediate does  exist. Methods such as the phi-value analysis have shown that the transition state (hereafter referred to as TS) of two-state globular proteins is structurally similar to the native state \cite{matouschek1989,baldwin2013,dijkstra2018}. The TS of two-state folders is a disguised molten-globule of very short lifetime whose thermodynamic properties reflect those of the molten-globule intermediate. In contrast to an intermediate state, defined as a local minimum in the free-energy landscape, the TS corresponds to a local maximum in the free-energy landscape.

A new direction of thought emerged in the late 80's by Wolynes and collaborators who proposed the energy landscape hypothesis (ELH): proteins fold in a funnel-like energy landscape by following different and productive folding trajectories \cite{frauenfelder1991,bryngelson1995}. Albeit not excluded intermediates are not obligatory folding steps. In both scenarios, MGH and ELH, the thermodynamics of the TS has generic and unique properties: on the one hand, a large energy barrier separates TS and N; on the other hand, there is a large configurational entropy loss upon forming the TS from the random coil or unfolded state (hereafter denoted as U). More recently, the alternative foldon hypothesis (FH) has gained considerable attention \cite{baldwin1995,maity2005}, based on the accumulated evidence gathered from hydrogen exchange, NMR, and mass spectrometry studies.  In the FH, proteins fold following a unique pathway by the cooperative and sequential formation of native structure domains (denoted as {\it foldons}). Folding amounts to the productive tinkering of amino acids and foldons rather than the diffusion of a polypeptide in a funnel-like energy landscape.

To evaluate the different hypotheses computer simulations and experiments are employed \cite{camacho1993,dill2012}. For the latter, it is crucial to have tools for accurately measuring the thermodynamics and kinetics of folding. Besides bulk techniques (e.g., NMR, mass spectrometry, calorimetry, etc.), single-molecule fluorescence and force spectroscopy offer complementary insights on the protein folding problem. With these, individual proteins are manipulated and monitored with enough temporal resolution to detect short-lived intermediates \cite{borgia2008,thirumalai2010,bustamante2020}, and measure transition path times along kinetic barriers \cite{cossio2018}. Key results are the demonstration that the ribosome promotes the efficient folding of the nascent polypeptide chain \cite{kaiser2011}, and the role of protein mechanical properties on nuclear translocation \cite{infante2019}. Single-molecule evidence of protein folding intermediates has been reported for RNAseH \cite{cecconi2005,stockmar2016}, the coiled-coil leucine zipper \cite{gebhardt2010,neupane2016} and  calmodulin \cite{stigler2011}. Recently, the molten-globule of apomyoglobin has been shown to be highly deformable under force \cite{elms2012} and an off-pathway molten-globule has been observed in apoflavodoxin \cite{lindhoud2015}. In other cases, proteins fold in a two-states manner without detectable intermediates (e.g., PrP protein \cite{neupane2016}), and a molten-globule of very short lifetime transiently forms along the folding pathway \cite{sosnick1994,fersht_1995}.

\begin{figure*}
\centering
\includegraphics[]{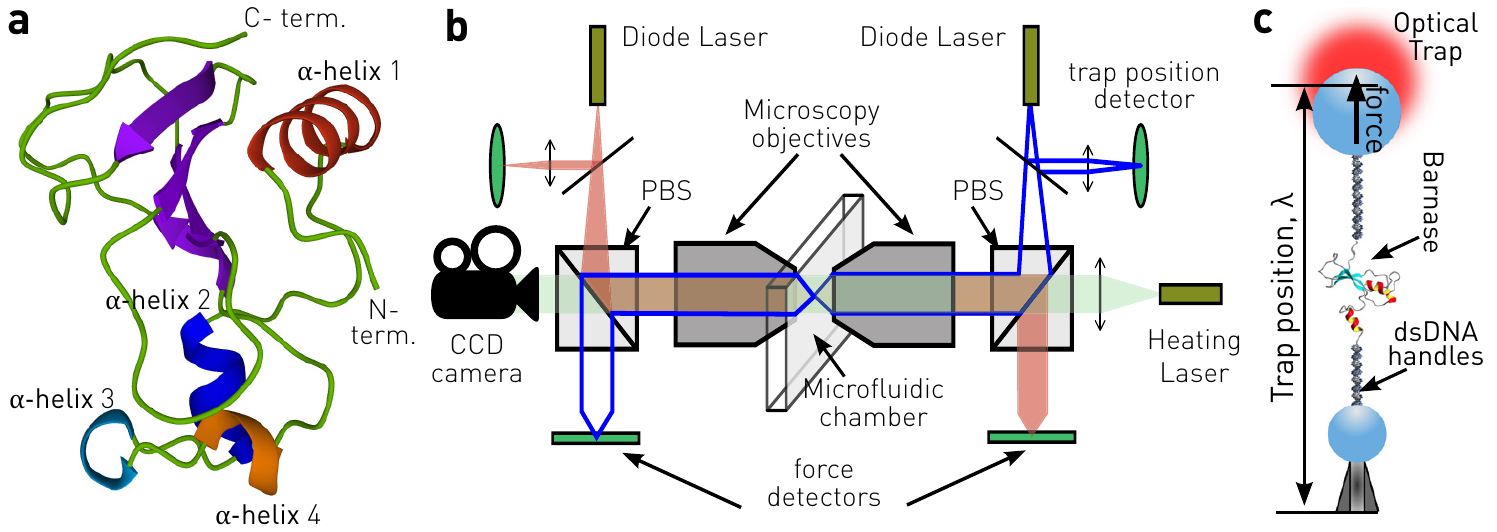}
\caption{Calorimetric force spectroscopy of protein barnase. \textbf{(a)} 3D view of native barnase obtained with X-ray diffraction with 1.50 $\textup{\r{A}}$ resolution \cite{martin1999}. Four external $\alpha$-helices (helix 1: Phe7-Tyr17 (red); helix 2: Lys27-Leu33 (blue); helix 3: Ala37-Lys39 (cyan); helix 4: Leu42-Val45 (orange)) contain a total of 25 amino acids surrounding four $\beta$-strands (purple) located in the protein core.  \textbf{(b)} Schematics of the temperature-jump optical trap setup. The diode lasers (red and blue) form a single optical trap, while the collimated heating laser (green) passes through the microfluidic chamber.  \textbf{(c)} Illustration of the molecular construct and experimental setup: barnase is flanked by two identical 500bp dsDNA handles and tethered between two beads. One bead is captured in the optical trap while the other one is kept fixed by air suction on the tip of a glass micro-pipette.}
\label{fig1}
\end{figure*}

Over the past decades, there has been much effort in determining the thermodynamic properties of intermediate and transition states in globular proteins. How much the enthalpy and entropy of the TS do differ from the native state? What is the heat capacity change ($\Delta C_p$) between the TS and the N and U conformations? How the TS properties change by varying the external conditions (e.g., temperature, ionic strength, and pH)? Answering these questions is essential to understand the features of the different hypotheses (e.g., the {\it liquid-like} properties of the molten-globule in the MGH or the funnel's shape in the ELH), and the nature of the folding process itself. 

Upon heating, proteins melt at a characteristic temperature $T_m$ at which the heat capacity at constant pressure, $C_p$, shows a peak \cite{ATikhomirova_2004,CJohnson_2013}. The heat capacity change upon folding, $\Delta C_p$, can be used to determine temperature-dependent enthalpy and entropy differences \cite{myers1995, gomez199}. Moreover, $\Delta C_p$ is directly related to the change in the number of degrees of freedom, $\Delta n$, across the transition, $\Delta C_p= \Delta n \cdot k_B/2$. Therefore $\Delta C_p$ quantifies the configurational entropy loss, the main contribution to the folding entropic barrier.   

Laser optical tweezers \cite{smith2003} are suitable for calorimetric measurements, however most studies have been carried out at ambient temperature $T_{\rm room}=298K$  \cite{tinoco_2002,keller_2003,ritort_2006}, due to the difficulty of controlling temperature \cite{williams2001entropy,stephenson2014combining}. For many years this limitation has challenged direct enthalpy and entropy measurements over a wide range of temperatures, rendering $\Delta C_p$ inaccessible to single molecule assays. We have recently built a temperature-jump optical trap suitable for single-molecule force spectroscopy above and below $T_{\rm room}$, thus providing a new calorimetric force spectroscopy (CFS) tool for molecular thermodynamics \cite{SLorenzo_2015,MRico18}. 
Here, we investigate the folding thermodynamics and kinetics of protein barnase, a paradigmatic model in protein folding studies (Fig.\ref{fig1}a). Barnase is a 110 amino acids bacterial ribonuclease globular protein secreted by \textit{Bacillus amyloliquefaciens}, which in physiological conditions degrades RNA in the absence of its protein inhibitor barstar \cite{VMitkevich_2003}. The high solubility and stability of barnase makes it an excellent model to investigate the folding kinetics of globular proteins, by combining phenomenological approaches (e.g., the phi-value analysis) with protein engineering and site-directed mutagenesis methods \cite{matouschek1989}. Barnase reversibly folds in a two-states manner between the unfolded and native conformations. It has been suggested  that barnase folds via a short-lived intermediate \cite{MBycroft_1990,fersht_1995,RBest_2001} and two transition states \cite{salvatella2005}. However pulling experiments at room temperature could not find evidence of intermediates down to milliseconds \cite{AAlemany_2016}. Here, we pull barnase in the range 7-37ºC and derive thermodynamic quantities by combining fluctuation theorems for free energy prediction and kinetics. We determine the temperature-dependent folding free energy ($\Delta G$), entropy ($\Delta S$), and enthalpy ($\Delta H$), to derive $\Delta C_p$ ($\sim 1000$ cal/mol$\cdot$K). Our results are consistent with calorimetry studies under similar ionic strength and pH conditions. 

We also determine the entropy, enthalpy and $\Delta C_p$ of the TS, finding that it is structurally similar to N. Upon folding, most of the enthalpy and entropy change occurs between N and TS, where roughly 80$\%$ of the native bonds are formed (from molten to native). In contrast, most of the folding $\Delta C_p$ occurs between U and TS, with $\sim$90$\%$ of configurational entropy loss. The collapse from TS to N mostly contributes to the enthalpy and entropy of folding, but residually to $\Delta C_p$. Our results demonstrate that the TS has the properties of a molten-globule: a large entropy and enthalpy relative to the native state and a low configurational entropy. Albeit structurally similar to the native, the molten-globule is a high energy state with most bonds not formed.

\begin{figure*}[ht]
    \centering
    \includegraphics{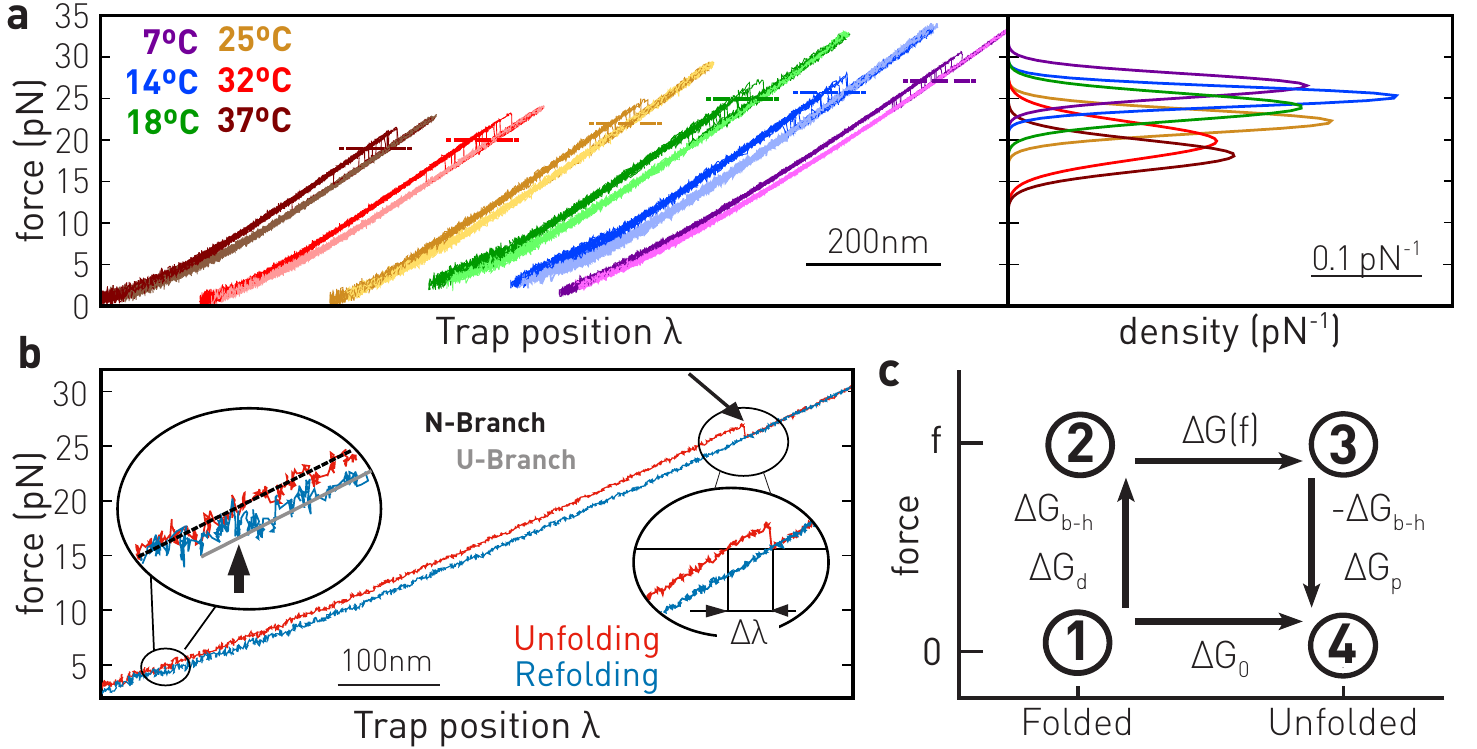}
    \caption{Pulling experiments of protein barnase. \textbf{(a)} Left: Unfolding (dark) and folding (light) force-distance curves (FDCs) at different temperatures, 7ºC (purple), 14ºC (blue), 18ºC (green), 25ºC (yellow), 32ºC (red) and 37ºC (brown). FDCs at each temperature have been shifted along the x-axis for clarity. Horizontal dashed lines denote the most probable unfolding force at each temperature. Right: Unfolding force distributions for all the studied temperatures. For sake of clarity, we show Gaussian fits to histograms. For higher temperatures less force is required to unzip the protein. \textbf{(b)} Unfolding (red) and refolding (blue) trajectories at 25ºC. Left Zoom: Refolding event (arrow) from the U-branch (gray solid) to the N-branch (black dashed). Right Zoom: Unfolding event (arrow) highlighting $\Delta\lambda$. \textbf{(c)} Scheme of the different thermodynamic steps to measure $\Delta G_0$. Stretching of the folded protein (1 $\rightarrow$ 2), unfolding at a given force (2 $\rightarrow$ 3), releasing of the unfolded protein (3 $\rightarrow$ 4), and unfolding at zero force (1 $\rightarrow$ 4).}
    \label{fig2}
\end{figure*}

\section{Results}
\subsection{Force-distance curves} \label{FDC}
Barnase was inserted in a molecular construct that was tethered between two beads and mechanically pulled with a temperature-jump optical trap (Fig.\ref{fig1}b,c and Materials and Methods \ref{met:SET}). Force-distance curves (FDCs) were measured by repeatedly pulling barnase between minimum and maximum force values at different temperatures. Figure \ref{fig2}a shows FDCs of five selected pulling cycles (unfolding and refolding) for the six investigated temperatures (7, 14, 18, 25, 32, 37ºC). It is apparent that the lower the temperature, the higher the unfolding force and the FDC hysteresis. In Figure \ref{fig2}b we show a single pulling cycle at 25ºC. During stretching (red curve), barnase unfolding is observed as a sudden force rip ($\Delta f\sim 2$pN) in the FDC at forces $\sim 15-30$pN. Upon force release (blue curve), a folding transition is detected as a small force jump ($\sim 0.5$pN) at forces $\lesssim 5$pN. The left encircled zoomed inset shows the two force branches where barnase is folded (N-branch, black dashed line) and unfolded (U-branch, grey solid line). N and U-branches describe the elastic response of the molecular construct where barnase is in N and U, respectively. The relative trap position ($\lambda$) in the two branches contains the trap bead displacement plus the handles extension and the molecular extension. The difference between both branches at a given force, $\Delta\lambda$ (right circled zoomed inset), is the difference of molecular extensions between the polypeptide chain and the projection on the force axis of the dipole formed by the N- and C- termini of barnase.
%

\subsection{Folding free energy, entropy and enthalpy}
\label{thermodynamics}
Here we describe how to extract the temperature-dependent folding free energy of barnase at zero force, $\Delta G_0(T)$, from the measured free energy difference $\Delta G(T)$. $\Delta G_0=G_{\rm U}-G_{\rm N}$ equals the (positive) free energy difference between the native conformation (N) and the random coil state (unfolded, U). The nomenclature for free energy differences employed throughout the paper is the standard one in single-molecule and calorimetric bulk studies. $\Delta G_0$ can be measured in bulk assays at zero force, whereas CFS experiments measure free energy differences at a force,  $\Delta G(f)$. To derive $\Delta G_0$ from  $\Delta G(f)$ it is necessary to subtract contributions coming from the experimental setup, such as the displacement of the bead from the center of the optical trap and the stretching of the handles and the polypeptide chain \cite{severino2019}. The procedure is illustrated in Fig.\ref{fig2}c where the different stretching contributions correspond to free energy differences measured over three distinct steps ($1\rightarrow 2;2\rightarrow 3;3\rightarrow 4$). $\Delta G_0(T)$ can be derived (Materials and Methods, \ref{met:ENE}) from the measured $\Delta\lambda(f,T)$ and the coexistence force in equilibrium, $f_c(T)$ (defined by $G_{\rm N}=G_{\rm U}$ or $\Delta G(f_c(T))=0$):
\begin{equation}
    \Delta G_0(T)=\int_0^{f_c(T)}\Delta\lambda(f,T)df \ \rm .
    \label{eqGint1}
\end{equation}
From Eq.\ref{eqGint1} we derive the folding entropy and enthalpy, $\Delta S_0 = -\partial \Delta G_0 / \partial T$ and $\Delta H_0=\Delta G_0-T\Delta S_0$. For the entropy we find, 
\begin{equation}
    \Delta S_0(T) = -\frac{\partial f_c(T)}{\partial T}\Delta\lambda(f_c(T)) - \int_{0}^{f_c(T)}\frac{\partial \Delta \lambda(f',T)}{\partial T} df' \ \ .
    \label{eq:SdAldT}
\end{equation}

\begin{figure*}[ht]
    \centering
    \includegraphics[]{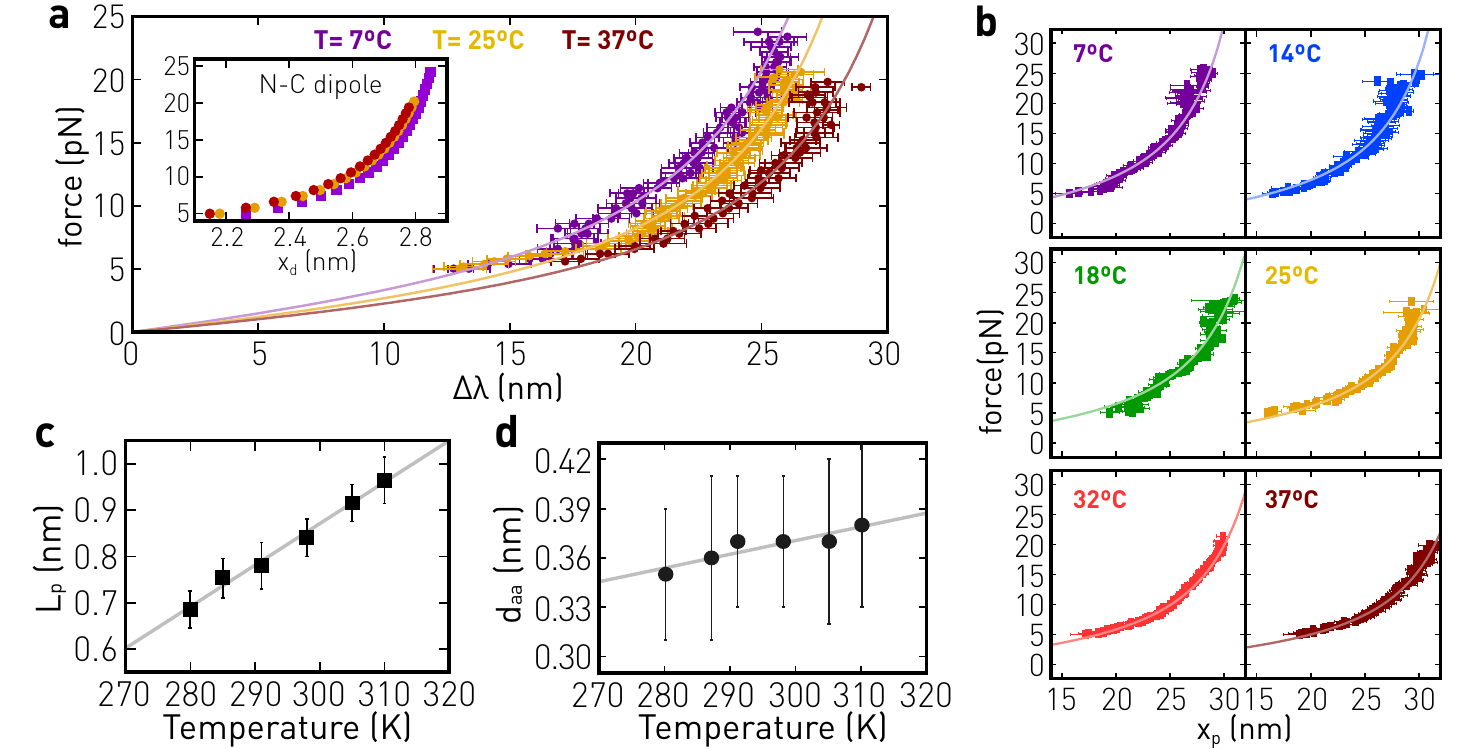}
    \caption{Elastic response of the polypeptide chain. \textbf{(a)} Force versus difference in trap position ($\Delta\lambda$) at three temperatures: 7ºC (purple), 25ºC (yellow) and 37ºC (brown). Inset: Elastic response of the folded protein modelled as a dipole of 3nm with the FJC model.\textbf{ (b)} Fits of the measured force versus  polypeptide chain extension ($x_p$) to the WLC model (solid lines). \textbf{(c,d)} Persistence length ($L_p$) and amino acid distance ($d_{\rm aa}$) of the polypeptide chain calculated from the fits in panel \textbf{b}. Solid lines are linear fits to the experimental points.}
    \label{fig3}
\end{figure*}

This is analogous to the Clausius-Clapeyron equation for first-order phase transitions, where $f$ and $\lambda$ are the equivalent of pressure and volume \cite{SLorenzo_2015}. The second term in Eq.\ref{eq:SdAldT} is the entropic contribution to stretch and orient the protein-dipole from zero force to $f_c$. Equations \ref{eqGint1} and \ref{eq:SdAldT} require measuring $\Delta\lambda(f,T)$ over the full integration range, [0, $f_c(T)$], and $f_c(T)$. While $\Delta \lambda$ is directly obtained from the FDCs (Fig.\ref{fig2}b, right zoom) the value of $f_c(T)$ is unknown due to the strongly irreversible FDCs. $f_c(T)$ might be extracted from equilibrium hopping experiments, however this is not possible in barnase due to the exceedingly long 
hooping times \cite{gebhardt2010,neupane2016}. Here we derive $f_c(T)$ by measuring $\Delta G_0(T)$ using the fluctuation theorem (Sec. \ref{thermo}) and using Eq.\ref{eqGint1}.

\subsection{Elastic response of the polypeptide chain}
\label{elastic}
The temperature-dependent elastic properties of the polypeptide chain were determined using the molecular extension $x_{\rm p}(f,T)$ obtained from $\Delta\lambda(f,T)$ (Fig.\ref{fig2}b) and the dipole contribution $x_{\rm d}(f,T)$ from $x_{\rm p}=\Delta\lambda+x_{\rm d}$. In Figure \ref{fig3}a we show $f$ versus $\Delta\lambda$ measured at three selected temperatures (7, 25 and 37ºC). To extract  $x_{\rm p}(f,T)$ from Eq.\ref{eqAL} we modeled the dipole extension $x_{\rm d}(f,T)$ with the Freely-Jointed Chain elastic model (Fig.\ref{fig3}a, inset), assuming that the distance between the N- and C-termini for the folded barnase (the dipole length taken equal to 3nm) is constant with temperature. By comparing Figure \ref{fig3}a (inset and main) we observe that $x_{\rm d}(f,T)\ll \Delta \lambda(f,T)$, as expected since the dipole length is much shorter than the polypeptide extension. Therefore, $x_{\rm p}(f,T)$ increases with $T$ at a given $f$, making the polypeptide chain stiffer with temperature. $x_{\rm p}(f,T)$ is well described by the inextensible Worm-Like Chain (WLC) model and its interpolation formula \cite{Siggia_1994},
\begin{equation}
    f = \frac{k_BT}{4L_{\rm p}}\left( \left( 1-\frac{x_{\rm p}(T)}{N_{\rm aa}\cdot d_{\rm aa}}\right)^{-2} + 4\cdot \frac{x_{\rm p}(T)}{N_{\rm aa}\cdot d_{\rm aa}} -1   \right)
\label{eq:WLC}
\end{equation}
where $L_{\rm p}$ is the persistence length, $N_{\rm aa}$ is the number of residues (110 for barnase), and $d_{\rm aa}$ is the distance between consecutive amino acids. The data relative to each investigated temperature were fit to Eq.\ref{eq:WLC}, as shown in Fig.\ref{fig3}b, with $L_p$ and $d_{\rm aa}$ free parameters. As reported in Figure \ref{fig3}c, $L_p$ shows a strong $T$ dependence, which is well approximated by a linear function of slope $0.011\pm0.001$ nm/K. Similar results are obtained if we fit the data with a perturbative expansion of the WLC model \cite{CBouchiat_1999} instead of Eq.\ref{eq:WLC}. Moreover, $d_{\rm aa}$ presents a weak $T$-linear dependence of slope $0.0008\pm0.0002$ nm/K (Fig.\ref{fig3}d), which is one order of magnitude smaller than for $L_{\rm p}$. Therefore, $d_{\rm aa}$ can be taken as constant, $\sim 0.37$ nm. Both fitting parameters at room temperature (298K) agree with previous results \cite{AAlemany_2016,RBest_2001}. 

\begin{figure*}[ht]
    \centering
    \includegraphics[]{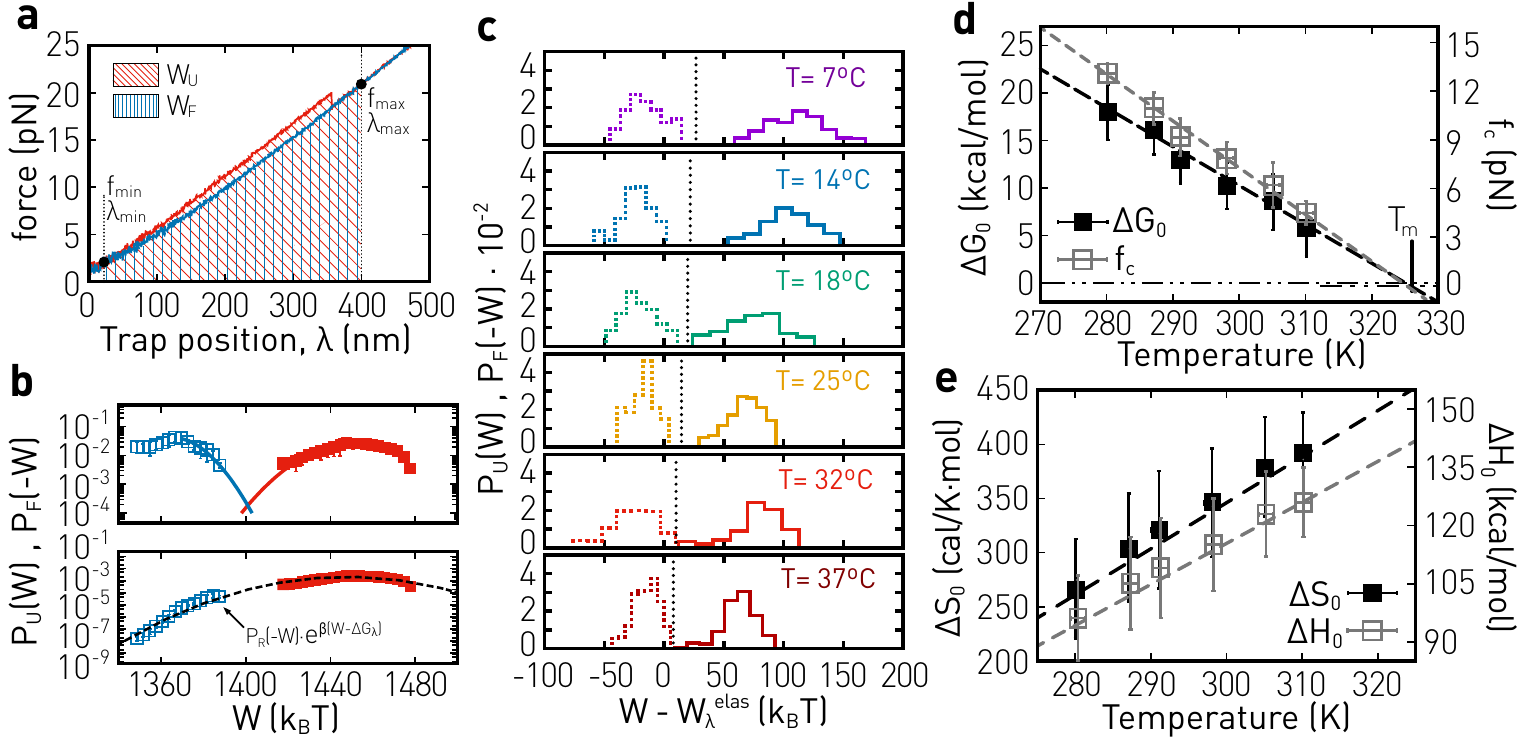}
    \caption{Folding thermodynamics of barnase. \textbf{(a)} Work measurements from unfolding (red) and folding (blue) FDC. The unfolding and folding work $W$  is the area below the FDC limited by $\lambda_{min}$ and $\lambda_{max}$ (red and blue areas). \textbf{(b)} Top: Unfolding (red full squares) and folding (blue empty squares) work distributions at 25ºC. Bottom: $P_U(W)$ and $P_F(-W)$ times $\exp((W-\Delta G_{\lambda})/k_BT)$. The black dashed line is a Gaussian fit to determine $\Delta G_{\lambda}$. \textbf{(c)}  Unfolding (solid line) and folding (dashed lines) work distributions at different temperatures (with the elastic contribution $W_{\lambda}^{\rm elas}$ being subtracted from the total work). The dotted vertical line indicates $\Delta G_0$ for each temperature. \textbf{(d)} Solid squares are the estimated $\Delta G_0$ versus $T$ and black dashed-line is a fit to $\Delta G_0(T) = \Delta H^m_0 - T\Delta S^m_0$. Empty squares are the estimated coexistence force and dashed gray line is a linear fit to $f_c$. \textbf{(e)} Solid and empty squares are the entropy and enthalpy differences at zero force. Dashed lines are fits to Eq.\ref{eq:AS_AH}a and Eq.\ref{eq:AS_AH}b, respectively.}
    \label{fig4}
\end{figure*}
\subsection{Measurement of folding free energy}
\label{thermo}
To derive $\Delta G_0$ we use the thermodynamic relation illustrated in Fig.\ref{fig2}c, that relates $\Delta G_0$ with the elastic contributions of the polypeptide chain $\Delta G_{\rm p}$, the dipole term $\Delta G_{\rm d}$, and $\Delta G(f)$ (Eq.\ref{eqG} in Materials and Methods \ref{met:ENE}), 
\begin{equation}
    \Delta G_0 = \Delta G_{\rm d}(0\to f)+\Delta G(f) -\Delta G_{\rm p}(0\to f) 
\label{eqG2}
\end{equation}
From the results of the previous section we can readily determine $\Delta G_{\rm p}$ and $\Delta G_{\rm d}$, however it remains unanswered how to measure $\Delta G(f)$. In optical tweezers experiments, the relative trap position $\lambda$ is the control parameter, rather than the force which fluctuates depending on the molecular state. A thermodynamic relation similar to Eq.\ref{eqG2} holds by a Legendre transforming $f\to\lambda$ to the $\lambda$-ensemble (section S1, Supp. Info.). $\Delta G_0$ is determined by measuring the free energy difference, $\Delta G_\lambda$, between minimum and maximum trap positions where barnase is folded ($\lambda_{\rm min}$) and unfolded ($\lambda_{\rm max}$), 
\begin{equation}
    \Delta G_0 = \Delta G_\lambda-W_{\lambda}^{\rm elas}
    \label{eq:AG00}
\end{equation}
where $W_{\lambda}^{\rm elas}$ stands for the elastic contributions of the setup (bead, handles, polypeptide chain and protein dipole) that must be subtracted to $\Delta G_\lambda$ (section S1, Supp. Info.).  

We used the work fluctuation theorem (work-FT) \cite{collin2005,jarzynski2011} to determine $\Delta G_\lambda$ from irreversible work ($W$) measurements by integrating the FDC between the two selected trap positions, $W = \int _{\lambda_{min}}^{\lambda_{max}} f d\lambda$ (Fig.\ref{fig4}a). Let $P_U(W)$ and $P_F(W)$ denote the unfolding and folding work distributions measured over many pulling cycles. The work-FT is given by,  
\begin{equation}
    \frac{P_U(W)}{P_F(-W)}=\exp{\Bigl( \frac{W-\Delta G_\lambda}{k_BT}\Bigr)}
    \label{ft}
\end{equation}
where $\Delta G_\lambda$ equals the reversible work. The minus work sign in $P_F(-W)$ in the left-hand-side of Eq.\ref{ft} is a consequence of the fact that $W<0$ in the folding process. A corollary of Eq.\ref{ft} is the Jarzynski equality, $\Delta G_\lambda = -k_BT \,\log \langle \exp (-W/k_BT) \rangle$, where $\langle (...)\rangle$ is the average over many (infinite) realizations. In practice, the number of pulls is finite and the Jarzynski equality is strongly biased \cite{palassini_2011}. From Eq.\ref{ft}, work distributions cross at $W=\Delta G_\lambda$, i.e., $P_U(\Delta G_\lambda)=P_F(-\Delta G_\lambda)$. However, the crossing point is not observed due to the hysteresis of the FDCs, quantified by the area enclosed between the unfolding and folding FDC, (Fig.\ref{fig4}a). The dissipated work is in the range 50-100$k_BT$, and much larger than the value of $\Delta G_0$ (see below). 
In the absence of crossing one can use the matching method \cite{collin2005}, that gives reasonable free energy estimates and is simpler than other mathematical approaches \cite{shirts2003,palassini_2011}. In this method, the value of $\Delta G_\lambda$ is determined by matching the functions $P_U(W)$ and $P_F(-W) \exp( (W-\Delta G_\lambda)/k_BT)$. In practice, the leftmost (rightmost) tails of $P_U(W)$ ($P_F(-W)$) are fitted to the generic form, $\sim\exp\left( -|W-W_{\rm max}|^\delta /\Omega\right)$, to extract the values of $\delta,\Omega,W_{\rm max}$ \cite{palassini_2011}. In Figure \ref{fig4}b (top) we show $P_U(W)$, $P_F(-W)$ at 25ºC and the fitted tails. Rightmost ($P_F$) and leftmost ($P_U$) tails are well fitted with $\delta \sim 1.9$, indicating Gaussian-like tails ($\delta = 2$). Therefore, we simultaneously fitted  $P_U(W)$ and $P_F(-W) \exp((W-\Delta G_\lambda)/k_BT)$ to a single Gaussian distribution (black dashed line in Fig.\ref{fig4}b bottom) to find the best matching the value of $\Delta G_\lambda$. The fact the generic and Gaussian distributions are nearly the same and a single Gaussian distribution (dashed line in Fig.\ref{fig4}b bottom) simultaneously fits $P_U(W)$ and $P_F(-W) \exp(W-\Delta G_\lambda)/k_BT)$ 
demonstrates that $\delta \sim$ 2 is an excellent approximation to $P_U(W)$ and $P_F(-W)$ tails around $W=\Delta G_\lambda$ (Fig. S1 in Supp. Info.).

To determine $\Delta G_0$ from $\Delta G_\lambda$ in  Eq.\ref{eq:AG00}, we subtract the elastic contributions as follows: the bead contribution was calculated by modeling the optical trap with Hooke's law, i.e., a constant optical trap's stiffness equal to 0.07pN/nm throughout the explored force range  \cite{NForns_2011,severino2019}; the DNA handles term was calculated by integrating the WLC model with the temperature-dependent elastic parameters from \cite{geggier_2010}; $\Delta G_{\rm p}$ and $\Delta G_{\rm d}$ contributions were calculated using the elastic parameters from section \ref{elastic}. In Figure \ref{fig4}c we show the $P_U(W)$ and $P_F(-W)$ at different temperatures. Distributions are plotted versus $W-W_{\lambda}^{\rm elas}$ instead of $W$, to directly determine $\Delta G_0$ with the matching method (Fig.\ref{fig4}c, dotted vertical lines). The values of $\Delta G_0(T)$ present a clear temperature dependence (Fig.\ref{fig4}d, filled black squares) as expected from the relation, $\Delta G_0 = \Delta H_0 - T \Delta S_0$. 

\subsubsection{Derivation of entropy and enthalpy}
\label{derivfc}
From Eq.\ref{eqGint1} and the measured values of $\Delta G_0(T)$ and $\Delta \lambda(f,T)$ (Sec. \ref{elastic}) we derive $f_c(T)$ in the range 7-37ºC (gray empty squares in Fig.\ref{fig4}d). $f_c(T)$ decreases linearly with $T$, thus defining a $f-T$ phase diagram separating the native and unfolded states. We note that the linear trend observed in  Fig.\ref{fig4}d does not agree with predictions by lattice models \cite{klimov1999, hyeon2005, imparato2009} for the critical force at which the fraction of native contacts equals 0.5 (Fig. S3 and Fig. S4 in Supp. Info.). The line $f_c(T)$ crosses the $T$-axis at $T_m\simeq 50$ºC, in agreement with bulk experiments (see below). Finally, from Eq.\ref{eq:SdAldT} and $f_c(T)$ we derived $\Delta S_0(T)$ (black solid squares in Fig.\ref{fig4}e). It changes by roughly $22\%$ in the whole temperature range indicating a finite $\Delta C_p$. Notice that the numerical $T$-derivative of $\Delta G_0(T)$ is roughly constant (Fig.\ref{fig4}d), which confirms Eq.\ref{eq:SdAldT} as the most reliable way to estimate $\Delta S_0(T)$. Folding enthalpies, $\Delta H_0=\Delta G_0 + T \Delta S_0$, are shown in Fig.\ref{fig4}e (empty squares). 

\subsubsection{Heat capacity change}
\label{DCp}
Bulk assays have shown that barnase has a finite $\Delta C_p$. The marked temperature dependence in $\Delta S_0$ and $\Delta H_0$ (Fig.\ref{fig4}e) allows us to extract $\Delta C_p$  across the melting transition. To do so, we expand $\Delta H_0$ and $\Delta S_0$ around the melting temperature $T_m$,

\begin{subequations}\label{eq:AS_AH}
    \begin{align}
        \Delta S_0(T) = \Delta S_0^m + \Delta C_p \cdot \log \left( \frac{T}{T_m} \right) \\
         \Delta H_0(T) = \Delta H_0^m + \Delta C_p \cdot \left( T-T_m \right) ~,
    \end{align}
\end{subequations}
where $\Delta S_0^m$ and $\Delta H_0^m=T_m\Delta S_0^m$ are the entropy and enthalpy at $T_m$, and $\Delta C_p$ is the heat capacity change between N and U. $\Delta S_0(T)$ and $\Delta H_0(T)$ were fitted to Eqs.\ref{eq:AS_AH}a,\ref{eq:AS_AH}b (dashed lines in Fig.\ref{fig4}e) with $\Delta C_p$, $\Delta H_0^m$, $\Delta S_0^m$, and $T_m$ fitting parameters. We obtain $\Delta C_p = 1030 \pm43$ cal/mol$\cdot$K, $\Delta S_0^m = 431 \pm 10$ cal/mol$\cdot$K and $\Delta H_0^m = 140 \pm 6$ kcal/mol.

\begin{figure*}[ht]
    \centering
    \includegraphics[]{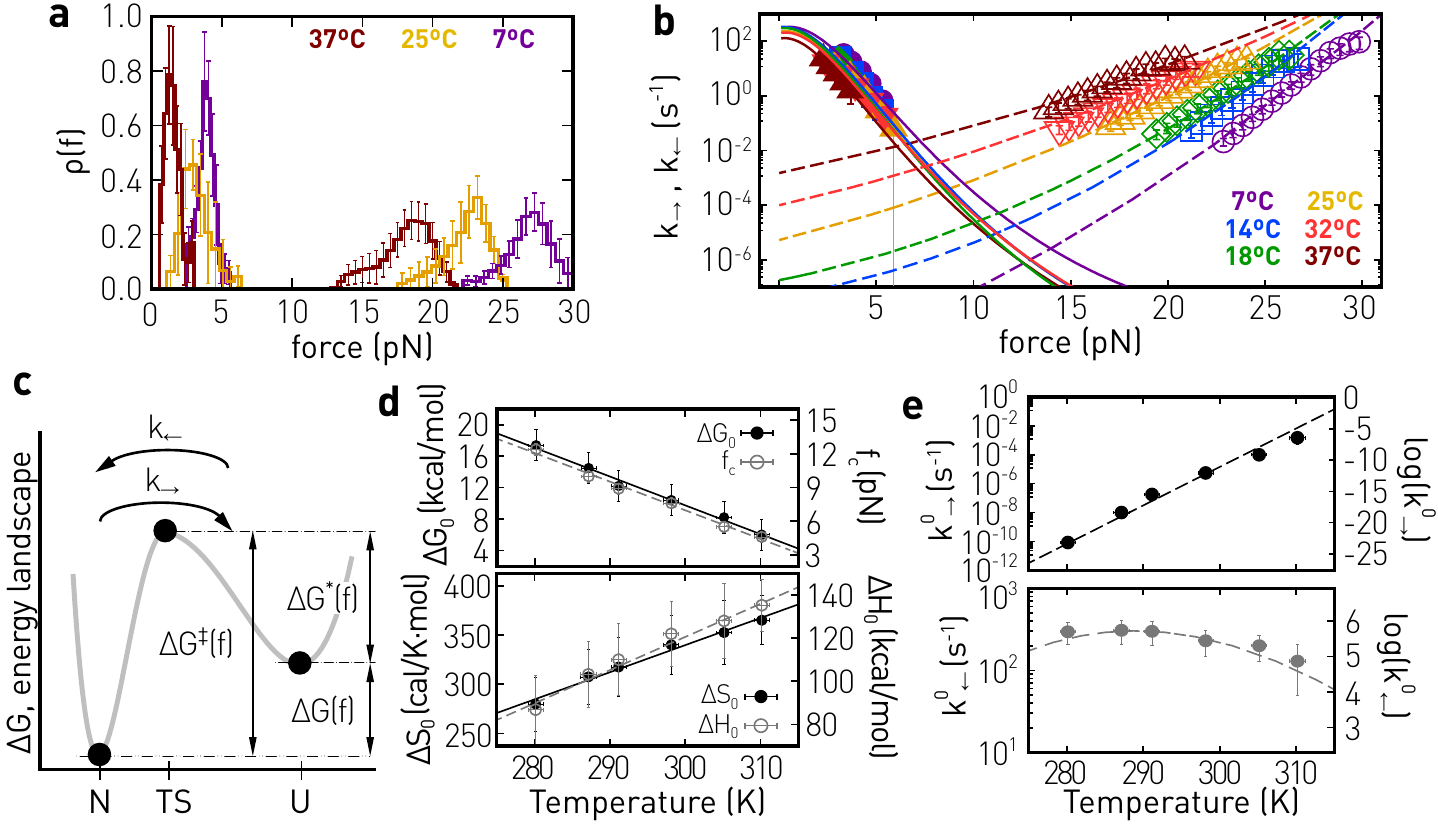}
    \caption{Folding  kinetics of barnase. \textbf{(a)} Unfolding (right) and refolding (left) force distributions at 7ºC (purple), 25ºC (yellow) and 37ºC (dark brown). \textbf{(b)} Force-dependent unfolding (empty symbols) and folding (solid symbols) kinetic rates for all temperatures, 7ºC (purple), 14ºC (blue), 18ºC (green), 25ºC (yellow), 32ºC (red), 37ºC (brown). The solid (refolding) and dashed (unfolding) lines are fits to the Bell-Evans model. \textbf{(c)} Schematics of the folding free energy landscape in the Bell-Evans model \textbf{(d)} Folding free energy and coexistence force (top) and entropy and enthalpy (bottom) as a function of temperature. Dashed lines are fits to Eqs.\ref{eq:AS_AH}a,b. \textbf{(e)} Unfolding (top) and refolding (bottom) kinetic rates at zero force versus $1/T$. Dashed lines are simultaneous fits to Eq.\ref{eq:k0_T}a,b. }
    \label{fig5}
\end{figure*}

\subsection{Kinetics}
\label{kinetics}

The results derived from the work-FT are confronted with those derived from the temperature-dependent unfolding and folding kinetic rates using the same experimental FDCs. Figure \ref{fig5}a shows the unfolding and folding force distributions ($\rho_{\rightarrow}(f)$, $\rho_{\leftarrow}(f)$) at three selected temperatures (7, 25 and 37ºC). We extract the unfolding and folding kinetic rates, $k_{\rightarrow}(f)$ and $k_{\leftarrow}(f)$, from the corresponding survival probabilities. If $f$ is ramped at constant loading rate $r=|df/dt|$, the following relations hold,
\begin{subequations}
    \begin{align}
        \frac{dP_{N}(f)}{df}= -\frac{k_{\rightarrow}(f)}{r}P_{N}(f) \,\, \Rightarrow \,\,
k_{\rightarrow}(f)=r\,\frac{\rho_{\rightarrow}(f)}{P_{N}(f)} \\
\frac{dP_{U}(f)}{df}= \frac{k_{\leftarrow}(f)}{r}P_{U}(f)\,\, \Rightarrow \,\,
k_{\leftarrow}(f)=r\,\frac{\rho_{\leftarrow}(f)}{P_{U}(f)}
    \end{align}
\end{subequations}
with $P_{N}(f)=1-\int_{0}^f\rho_{\rightarrow}(f)df$ and $P_{U}(f)=1-\int_f^{\infty}\rho_{\leftarrow}(f)df$ being the survival probabilities of N and U, respectively. Our measurements of $k_{\rightarrow}(f)$ and $k_{\leftarrow}(f)$ are shown in Figure \ref{fig5}b in a log-normal scale, for all the studied temperatures. The force dependence of the kinetic rates is described by,
\begin{subequations}\label{eq:kFU}
    \begin{align}
         k_{ \rightarrow }(f) = k_a\exp \left( -\frac{\Delta G^{\ddagger}(f)}{k_BT}\right) \\
         k_{ \leftarrow }(f) = k_a\exp \left( -\frac{\Delta G^{*}(f)}{k_BT}\right) ~,
    \end{align}
\end{subequations} 

with $k_a$ the attempt rate and $\Delta G^{\ddagger}(f)$ ($\Delta G^{*}(f)$) the force-dependent kinetic barrier relative to N (U). An illustrative free energy landscape is shown in Fig.\ref{fig5}c highlighting $\Delta G(f)$, $\Delta G^{\ddagger}(f)$, and $\Delta G^{*}(f)$. Notice the hysteresis between unfolding and folding (a minimum of $\sim$10pN gap is observed between the measured unfolding and folding rates, Fig. \ref{fig5}b). This fact precludes us from determining the coexistence force and the folding free energy at zero force using \eqref{eqGint1}. To circumvent this problem, we have determined the force dependence of the unfolding and folding kinetic rates beyond the Bell-Evans model, where distances of N and U to the TS are taken as force independent \cite{Hyeon2007,Rico2021}. To do so, we have used the detailed balance condition,
\begin{equation}
    \frac{k_{\leftarrow}(f)}{k_\to(f)} = \exp \left(\frac{\Delta G(f)}{k_BT}\right)
    \label{eq:DBC}
\end{equation}
that relates the unfolding/folding kinetic rates with the energy difference between N and U at force $f$, $\Delta G(f)$ (c.f. \eqref{eqG2}). For a given trial value of $\Delta G_0$, the energy $\Delta G(f)$ is calculated using \eqref{eq:DBC} with the elastic contributions determined in section \ref{elastic}. Then we used \eqref{eq:DBC} to reconstruct the unfolding (folding) kinetic rates in the region of forces where folding (unfolding) transition events are observed in the pulling experiments. To estimate the quality of the continuity of the rates between the two force regimes, we have fitted the unfolding kinetic rates (measured and reconstructed) to a single quadratic function in the log-normal plot (dashed lines in Fig. \ref{fig5}b). From these fits and \eqref{eq:DBC}, we inferred the folding kinetic rates (solid lines in Fig. \ref{fig5}b). This procedure has been repeated by varying $\Delta G_0$ in the range of 2-20 kcal/mol in steps of 0.2 kcal/mol. To determine the $\Delta G_0(T)$ that best fits data, we have minimized the $\chi^2$ between the predicted force-dependent kinetic rates and the experimental and reconstructed rates (Fig. S2 in Supp. Info.). The curvature of the reconstructed $\log k_\rightleftarrows$ vs force plots at low forces (Fig. \ref{fig5}b) has been also observed for the ligand binding rates to filamin protein \cite{Rognoni2012}. The derived values for $\Delta G_0(T)$ are shown as black circles in Fig.\ref{fig5}d-top. In addition, using the extrapolated kinetic rates (solid and dashed lines in Fig. \ref{fig5}b), we determined the coexistence force $f_c$ as the force value at which $k_{\rightarrow} = k_{\leftarrow}$ (empty gray circles in Fig. \ref{fig5}d-top).

From $\Delta \lambda(f,T)$, derived in Section \ref{elastic}, and the values of $f_c (T)$ and $\Delta G_0 (T)$ obtained from the kinetics analysis, we used Eq.\ref{eq:SdAldT} to calculate $\Delta S_0(T)$ and $\Delta H_0(T)$ (Fig. \ref{fig5}d-bottom, black circles and empty gray circles, respectively). These values agree with those obtained independently from the work- FT analysis (Fig. S3 in Supp. Info.). Fitting $\Delta S_0(T)$ and $\Delta H_0(T)$ to Eqs.\ref{eq:AS_AH}a,b gives: $\Delta C_p = 1100\pm60$ cal/mol$\cdot$K, $\Delta S_0^m = 437 \pm 8$ cal/mol$\cdot$K and $\Delta H_0^m = 141 \pm 3$ kcal/mol.

Furthermore, we determined the TS enthalpy and entropy differences, $\Delta S^\ddagger,\Delta H^\ddagger,\Delta S^*,\Delta H^*$ relative to states N,U (Fig. \ref{fig5}c). To this aim, we rewrite the kinetic rates in Eqs.\ref{eq:kFU}a,b at zero force, $k_{\leftarrow}^0$, $k_{\rightarrow}^0$, in terms of the TS entropies and enthalpies,
\begin{subequations}\label{eq:k0_T}
    \begin{align}
         k_\rightarrow^0(T) = k_a \exp{\Bigl( \frac{\Delta S^{\ddagger}}{k_B}\Bigr )} \exp{\Bigl( -\frac{\Delta H^{\ddagger}}{k_BT}\Bigr )} \\
          k_\leftarrow^0(T) = k_a \exp{\Bigl( \frac{\Delta S^{*}}{k_B}\Bigr )} \exp{\Bigl( -\frac{\Delta H^{*}}{k_BT}\Bigr )}.
    \end{align}
\end{subequations}
We performed a simultaneous fit of $k_{\rightarrow}^0$ and $k_{\leftarrow}^0$ to Eq.\ref{eq:k0_T}a,b to derive the values of $\Delta S^{\ddagger},\Delta S^*$,$\Delta H^{\ddagger}$, $\Delta H^*$. Interestingly, we found that $k_{\rightarrow}^0$ is strongly $T$-dependent, while $k_{\leftarrow}^0$ depends weakly, hinting at an entropy-driven folding process. The four-parameters fit was done by imposing  two constraints: $\Delta S^* = \Delta S^{\ddagger} - \Delta S_0$; and $\Delta H^* = \Delta H^{\ddagger} - \Delta H_0$. For the fits to Eq.\ref{eq:k0_T}a,b the values of $\Delta S_0(T)$ and $\Delta H_0(T)$ have been taken as the mean values obtained from the FT (Fig.\ref{fig4}e) and kinetics (Fig.\ref{fig5}d-bottom). Moreover, we used the attempt rate previously obtained on the same molecular system in similar experimental conditions \cite{AAlemany_2016}, $k_a \sim 150 \rm s^{-1}$. Fits are shown as dashed lines in Fig. \ref{fig5}e. TS entropies and enthalpies are shown in Figure \ref{fig6}a and in Table \ref{tab:discusion} at all studied temperatures and at the average $T_m$ (50ºC). Fitting them to Eq.\ref{eq:AS_AH}a,b permits us to extract the heat capacity change between N and TS ($\Delta C_p^{\rm{N-TS}}$) and between TS and U ($\Delta C_p^{\rm{TS-U}}$). We obtain $\Delta C_p^{\rm{N-TS}}\sim 150$cal/mol$\cdot$K and $\Delta C_p^{\rm{TS-U}}\sim 900$cal/mol$\cdot$K, which gives the folding $\Delta C_p \sim 1050$cal/mol$\cdot$K.


\begin{table*}[bh]
    \centering
    \begin{tabular}{c||c|c|c||c|c|c||c|c|c}
          \textbf{$T$} & \textbf{$\Delta G_0$} & \textbf{$\Delta H_0$} & \textbf{$\Delta S_0$} & \textbf{$\Delta G^{\ddagger}$} & \textbf{$\Delta H^{\ddagger}$} & \textbf{$\Delta S^{\ddagger}$} & \textbf{$\Delta G^{*}$} & \textbf{$\Delta H^{*}$} & \textbf{$\Delta S^{*}$}    \\\hline
        
    7  & $17 \pm 3$ & $94 \pm 11$ & $273 \pm 40$ & $12\pm 4$ & $101\pm2$  & $318\pm 14$ & $-6\pm 4$  & $6\pm 3$   & $45\pm 15$  \\
    14 & $15 \pm 2$ & $104 \pm 12$ & $310 \pm 40$ & $11\pm 4$ & $103\pm2$  & $324\pm 14$ & $-4\pm 3$  & $0\pm 2$   & $14\pm 11$  \\
    18 & $13 \pm 2$ & $108 \pm 13$ & $327 \pm 43$ & $9\pm 4$  & $104\pm2$  & $326\pm 14$ & $-3\pm 3$  & $-3\pm 2$  & $-2\pm 10$  \\
    25 & $10 \pm 2$ & $116 \pm 12$ & $353 \pm 38$ & $8\pm 5$  & $105\pm2$  & $328\pm 15$ & $-3\pm 3$  & $-10\pm 2$ & $-25\pm 11$ \\
    32 & $8 \pm 2$ & $122 \pm 12$ & $372 \pm 39$ & $6\pm 4$  & $106\pm2$  & $329\pm 13$ & $-3 \pm 3$ & $-16\pm 2$ & $-42\pm 11$ \\
    37 & $6 \pm 2$ & $126 \pm 10$ & $387 \pm 29$ & $4\pm 3$  & $108\pm 2$ & $332\pm 12$ & $-2\pm 3$  & $-19\pm 2$ & $-56\pm 10$ \\ \hline
    50 & 0 &  $140 \pm 2$  & $434 \pm 10$ &  $2 \pm 2$ & $111 \pm 2$ & $337 \pm 4$ & $2 \pm 3$ & $-30 \pm 2$ & $-100 \pm 8$ \\  \hline
        \end{tabular}
    \caption{Thermodynamic properties of barnase at three selected temperatures: $T$ in ºC; $\Delta G$ and $\Delta H$ in kcal/mol; $\Delta S$ in cal/mol$\cdot$K. Thermodynamic potentials differences: $(_0)$, N-U; $(^{\ddagger})$ N-TS; $(^{*})$ U-TS. Note that $\Delta_{\rm N-U}=\Delta_{\rm N-TS} - \Delta_{\rm U-TS}$ or $\Delta_{0}=\Delta^{\ddagger}-\Delta^{*}$ in our notation.}
    \label{tab:discusion}
\end{table*}

\subsection{The folding funnel}
\label{funnel}
\begin{figure*}[ht]
    \centering
    \includegraphics[]{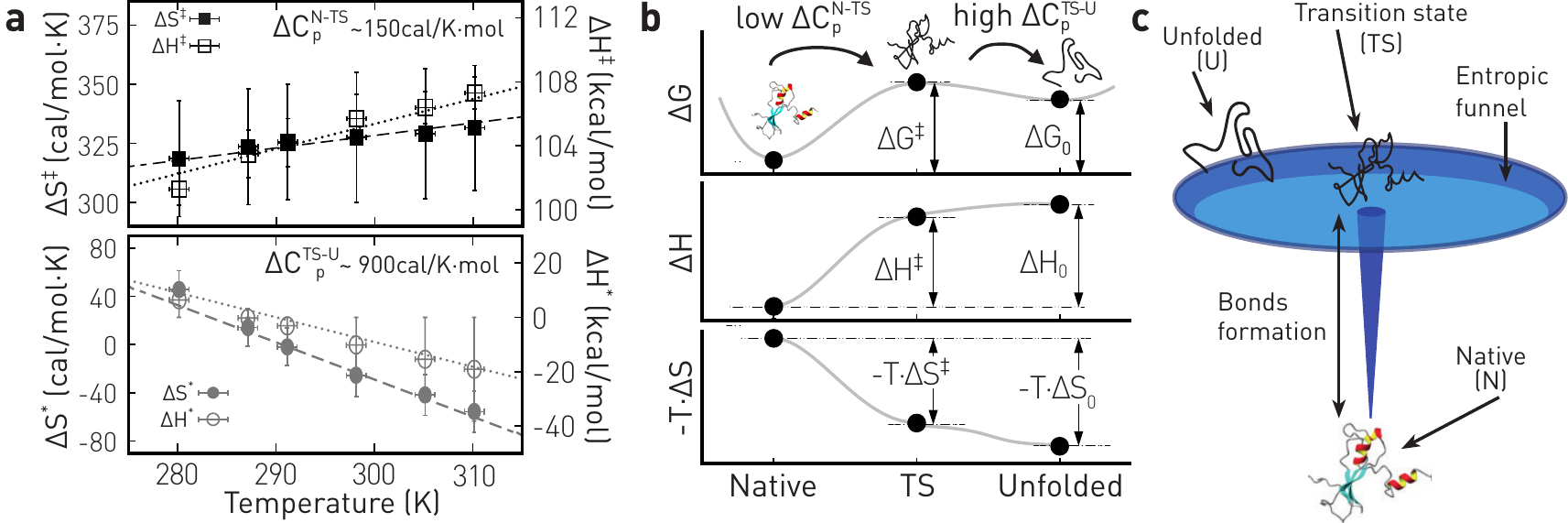}
    \caption{Barnase free energy landscape. \textbf{(a)} Entropy (full symbols) and enthalpy (empty symbols) differences between N and TS ($^\ddagger$, black); and between U and TS ($^*$, gray). \textbf{(b)} Illustration of the free energy ($\Delta G$, top), enthalpy ($\Delta H$, middle), and entropy ($-T\cdot\Delta S$, bottom) landscape relative to N. \textbf{(c)} Golf-hole course folding free energy landscape of barnase highlighting the molten-globule structure formed in the transition state (TS).}
    \label{fig6}
\end{figure*}

Figure \ref{fig6}a and Table \ref{tab:discusion} summarize our main results: the energy differences between states N and U ($\Delta G_0$, $\Delta H_0$ and $\Delta S_0$); the barrier energies to unfold, N-TS ($\Delta G^\ddagger$, $\Delta H^\ddagger$ and $\Delta S^\ddagger$); and the barrier energies to fold,  U-TS ($\Delta G^*$, $\Delta H^*$ and $\Delta S^*$). The energy parameters of the free energy landscape of barnase (Fig.\ref{fig5}c) are illustrated in Fig.\ref{fig6}b. Results show that barrier entropies, enthalpies and free energies to fold (U$\to$TS) are one order of magnitude smaller than the corresponding barriers to unfold (N$\to$TS): $|\Delta S^*| \ll |\Delta S^\ddagger|$, $|\Delta H^*| \ll |\Delta H^\ddagger|$ and $|\Delta G^*| \ll |\Delta G^\ddagger|$. This difference suggests a folding process in two steps (Fig.\ref{fig6}c). In a first step, the unfolded protein reaches a TS with a few H-bonds ($\sim 20\%$) formed relative to U. In a second step, the protein collapses into N by forming the rest of native bonds ($\sim 80\%$). These bond percentages are estimated from the different enthalpy values for the TS relative to N and U ($\Delta H^\ddagger \sim 109$ kcal/mol and $\Delta H^* \sim -30$ kcal/mol at $T_m$). 

A remarkable difference is found in $\Delta C_p$ between TS and N or U (Fig.\ref{fig6}a). The main contribution to the total $\Delta C_p= 1065 \pm 50$ cal/mol$\cdot$K comes from TS and U ($\Delta C_p^{\rm{TS-U}}= 905 \pm 20$ cal/mol$\cdot$K), which is $\sim$ 9 times larger than between N and TS ($\Delta C_p^{\rm{N-TS}}= 155 \pm 30$ cal/mol$\cdot$K) (Fig. S5, Supp. Info.). The value of $\Delta C_p$ is directly proportional to the change in the number of degrees of freedom ($\Delta n$), $\Delta C_p=\Delta n\cdot k_B/2$ which gives $\Delta n\sim1$ per cal/K$\cdot$mol unit in $\Delta C_p$. This gives $\Delta n^{\rm{TS-U}}\sim900\gg \Delta n^{\rm{N-TS}}\sim150$ showing that the main configurational entropy loss occurs upon forming the TS from U. This result depicts the TS as a molten-globule of high free energy ($\Delta G^\ddagger\sim \Delta G_0$) and low configurational entropy ($\Delta C_p^{\rm{N-TS}}\ll \Delta C_p$), which is  structurally similar to the native state: the major change in $\Delta C_p$ and $\Delta n$ occurs between U and TS.

\section*{Discussion}
We have used calorimetric optical tweezers to measure the FDCs of protein barnase in the range 7-37ºC and derived the folding thermodynamics at the single molecule level. An analysis based on the Clausius-Clapeyron equation (Eq.\ref{eq:SdAldT}) was used to extract the temperature-dependent values of $\Delta S_0(T)$ and $\Delta H_0(T)$ and $\Delta C_p$. These agree with those obtained from bulk experiments in our conditions of ionic strength (20mM monovalent salt) and neutral pH (7.0). The mean values for the enthalpy, entropy and $T_m$ derived from the work-FT and kinetic analysis ($\Delta H_0^m = 140 \pm 2$ kcal/mol, $\Delta S_0^m = 434 \pm 10$ cal/mol$\cdot$K and $T_m = 50\pm 2$ºC) agree with those reported in the literature and collected in  \cite{pfeil2012}, $\Delta H_m \sim 115 - 145$ kcal/mol and $\Delta S_m \sim 400$ cal/mol$\cdot$K (summarized in table S1 of Supp. Info.). Our estimation of $\Delta C_p = 1050 \pm 50$ cal/mol$\cdot$K also agrees with values obtained from differential scanning and isothermal titration calorimetry assays, $\Delta C_p \sim 1450 \pm 250$ cal/mol$\cdot$K (table S1, Supp. Info.), as well as with recent atomistic numerical simulations \cite{galano2019}. Measurements of $\Delta C_p$ in calorimetric experiments often require determining the temperature dependence of $\Delta H$ with pH, ionic strength, or the denaturant concentration. In contrast, with calorimetric force spectroscopy we directly measure thermodynamic potentials and kinetics at a given temperature. In Fig. S3 of Supp. Info., we also compare the protein stability curve of barnase derived from the fits to Eqs.\ref{eq:AS_AH}a,b with that reported in calorimetric studies \cite{YGriko_1994,makarov1993} and numerical simulations \cite{galano2019}.

Remarkably, entropies and enthalpies between the transition state, TS, and the unfolded state U, ($\Delta S^*$, $\Delta H^*$, $\Delta G^*$), are $\sim10$ times lower than the corresponding ones between TS and the native state N, ($\Delta S^\ddagger$, $\Delta H^\ddagger$, $\Delta G^\ddagger$). In fact, the low value of $\Delta G^*$ correlates with the high compliance of the molten-globule upon stretching, as has been shown for apomyoglobin \cite{elms2012}. A general feature of free energies ($0,\ddagger,*$) in Table \ref{tab:discusion} is the compensation observed between entropy ($T\Delta S$) and enthalpy ($\Delta H$) contributions, i.e., $\Delta G=\Delta H-T\Delta S\ll |\Delta H|,|T\Delta S|$. In contrast, the major contribution to $\Delta C_p$ occurs between U and TS ($\Delta C_p^{\rm{TS-U}}\gg \Delta C_p^{\rm{N-TS}}$). The value of $\Delta C_p^{\rm{TS-U}}$ is proportional to the reduction ($\Delta n^{\rm{TS-U}}\sim 900$) in the number of degrees of freedom (dof) between TS and U, at a rate of $\sim$ 1 dof/(cal/K$\cdot$mol). 

Our results agree with the folding funnel scenario of the energy landscape hypothesis (ELH) (Fig.\ref{fig6}c), where a large configurational entropy loss ($\sim 90\%$) occurs upon forming the TS from the unfolded state. Such entropy loss is accompanied by a low enthalpy change ($\sim 20\%$ of the total folding enthalpy). The large configurational entropy loss between U and TS demonstrates that folding is an entropically driven process in a golf-course energy landscape, where the TS is the native hole. The collapse from TS to N forms most of the native bonds accounting for most of the entropy and enthalpy of folding ($\Delta S_0\simeq \Delta S^\ddagger, \Delta H_0\simeq \Delta H^\ddagger$). Overall, our results also validate the main predictions of the ELH.

Are these results applicable to other proteins? Assuming an equal average enthalpy per native bond in N and TS, their structural similarity implies a low fraction of native bonds at TS, $f^\ddagger=\Delta H^*/ \Delta H_0$. For barnase, $f^\ddagger\sim$0.2 at $T_m$, a value that decreases as $T$ is lowered being $f^\ddagger\sim$0 at $T=7$ºC. These results are at odds with the observation often made in computational studies that $f^\ddagger$=0.5 at the TS. Previous AFM studies on the ddFLN4 protein domain in the range 5-37ºC show that $f^\ddagger$ increases with $T$, reaching a maximum ($f^\ddagger=0.19$) at 37ºC \cite{schlierf2005}. Apparently, the increase of TS stability with $T$ facilitates the collapse from TS to N.

The TS of barnase features the properties of a dry molten-globule: a native-like expanded structure with the backbone formed, but with side chains loosely packed \cite{baldwin2010}. The dry molten-globule has a large enthalpy relative to N ($\Delta H^\ddagger$) but a low $\Delta C_p^{\rm{N-TS}}$, the major contribution to $\Delta C_p$ being $\Delta C_p^{\rm{TS-U}}$, in agreement with our results. We hypothesize that folding proceeds in two steps. First (U to TS), hydrophobicity \cite{imai2007} drives the formation of the barnase backbone by the stabilization of the four $\alpha$-helices and $\beta$-strands and the expulsion of water from the protein core. The small difference between the net number of H-bonds between U and TS leads to small values of $\Delta H^*,\Delta S^*$ relative to $\Delta H_0,\Delta S_0$. Next (TS to N), the dry molten-globule collapses to N stabilized by the liquid-like Van der Waals interactions between the loosely packed side chains in TS. The $1/r^{6}$-dependence of Van der Waals interactions implies a large $\Delta H^\ddagger$ even for a short-distance collapse. $\Delta H^\ddagger$ is compensated by $T\Delta S^\ddagger$ (enthalpy-entropy compensation) a generic feature of weak interactions. Pushing the analogy further, protein folding resembles planet formation, where the mantle forms first and the core solidifies afterward.

Alternatively, in a wet molten-globule the backbone is formed at TS, but water remains inside the protein core. A wet molten-globule would lead to a high free energy barrier to folding, $\Delta G^*$, and a two-state folding scenario. Instead, $\Delta G^*=2\pm 3$ kcal/mol at $T_m$ and negative below $T_m$ (Tab. \ref{tab:discusion}). Remarkably, our results support the downhill folding (rather than two-states) scenario \cite{Eaton1999, Munoz2004, Liu2012} correlating it with the MGH.

Our results do not exclude the FH picture, where proteins fold along a well-defined pathway by the sequential formation of foldons. In the FH, the free energy landscape resembles the ELH golf course shown in Fig. \ref{fig6}c, except for the fact it incorporates a specific folding pathway (groove) along the ground with one or more intermediates (Fig. S6 in Supp. Info.). Discriminating between the ELH and FH would require monitoring folding paths in configurational space. Bulk measurements have addressed this question by combining hydrogen exchange \cite{Englander2017} and NMR \cite{Sadqi2006}. In single molecule experiments one might measure reaction coordinates other than the molecular extension, e.g., in multiple-color FRET \cite{Schuler2002}, atomistic simulations \cite{Lindorff2011, Kim2020} and changing environmental conditions and mutations \cite{Guinn2015}. Reproducible patterns in the folding trajectories are evidence of a preferential folding pathway in the energy landscape, supporting the FH.

Summing up, calorimetric force spectroscopy permits measuring folding entropies and enthalpies over a broad temperature range, with the accuracy necessary to determine heat capacity changes. In conjunction with a detailed kinetics study, this permits us to determine barrier entropies, enthalpies, and heat capacity changes relative to the native and unfolded states. Three thermodynamic inequalities summarize our results: $|\Delta S^*|\ll\Delta S^\ddagger,|\Delta H^*|\ll\Delta H^\ddagger$ and $\Delta C_p^{\rm{TS-U}} \gg \Delta C_p^{\rm{N-TS}}$. These are key inequalities for molecular folding in line with predictions of the molten-globule and energy landscape hypotheses. In particular, accurate measurements of $\Delta C_p$ are crucial to quantify to which extent configurational entropy loss drives intermediates formation and folding. Our study might be extended to other proteins, RNAs, and ligand-substrate binding \cite{Rognoni2012, Naqvi2015, Sonar2020}. In the latter, the ligand docks into the binding site of the substrate by searching in configurational space, similarly to finding the TS in protein folding. Docking is then followed by the assembly of the ligand-substrate complex, analogously to the TS-N collapse in protein folding.

\section*{Material and methods}
\subsection*{Molecular construct and experimental setup}\label{met:SET}
For pulling experiments barnase is expressed between two identical dsDNA (500bp) handles, which are attached to the N- and C-termini via cysteine-thiol chemical reduction (details in \cite{AAlemany_2016}). The 5'-end of one handle is labeled with a biotin, while the 3'-end of the other handle is labeled with a digoxigenin. The biotin and digoxigenin labeled ends specifically bind to spreptavidin (SA) and anti-digoxigenin (AD) coated beads. For the pulling experiments (Fig.\ref{fig1}c), one end is attached to the SA bead, which is kept fixed at the tip of a glass micro-pipette by air suction, whereas the other end is attached to the AD bead captured in the optical trap. Force changes by varying the relative distance $\lambda$ between the center of the optical trap and the bead in the pipette. In pulling experiments, the optical trap is moved up and down at a given speed and force ramped between an initial force ($\sim 1-2$pN), where the molecule is folded, and a maximum force ($\sim 30$pN), where barnase is unfolded. The unfolding transition is detected as a force rip in the force versus $\lambda$ curve (FDC). Moreover, due to folding reversibility of barnase, it refolds upon reducing $\lambda$ and the force. 

To perform CFS experiments, we used the temperature-jump optical trap described in \cite{SLorenzo_2015}. Briefly, a collimated laser at 1435nm wavelength (heating laser) is used to heat uniformly a $\sim $100$\mu$m radius area in the center of the fluidics chamber where the experiments are carried out. The wavelength is chosen to maximize the absorption by the water in the buffer solution (10mM of Na$_\text{2}$HPO$_\text{4}$ and NaH$_\text{2}$PO$_\text{4}$ at pH 7.0) to heat the surrounding medium to locally raise the temperature from 25ºC (room temperature) to 40ºC. Moreover, we can place the miniaturized optical tweezers instrument inside an icebox kept at 5ºC and heat from this basal temperature up to 25º using the heating laser. In this way, the available temperature ranges from 5º to 40ºC.

\subsection*{Folding free energy at zero force}\label{met:ENE}
To derive the folding free energy at zero force, $\Delta G_0$, from force experiments, we consider four different states of the protein (Fig.\ref{fig2}c): {\bf\circled{1}}  folded barnase at zero force; {\bf\circled{2}} folded barnase at a given force $f$; {\bf\circled{3}} unfolded barnase at the same force $f$; and {\bf\circled{4}} unfolded barnase at zero force. Notice that the direct unfolding pathway at zero force 1$\rightarrow$4, observed in bulk experiments, can be decomposed as the sum of three sequential steps (1$\rightarrow$4 = 1$\rightarrow$2 + 2$\rightarrow$3 + 3$\rightarrow$4). The three steps are as follows. {\it Step 1$\rightarrow$2:} folded barnase is reversibly pulled from zero force to force $f$ along the native branch of the FDC (left zoom in Fig.\ref{fig2}c, black dashed line). The free energy difference equals the sum of the reversible work to orient a dipole of length equal to the distance between the N- and C-termini of folded barnase ($\simeq$ 3nm \cite{AAlemany_2016}), $\Delta G_{\rm d}(0\to f)$, and the reversible work of stretching the handles and displacing the bead in the optical trap, $\Delta G_{\rm h-b}(0\to f)$. {\it Step 2$\rightarrow$3:} folded barnase is reversibly unfolded (denaturated) at a constant force $f$. In this step the free energy difference, $\Delta G(f)$, equals the free energy of the stretched polypeptide chain minus the folding free energy of native barnase, at force $f$. {\it Step 3$\rightarrow$4:} the stretched polypeptide chain is reversibly relaxed from $f$ to zero force along the unfolded branch of the FDC (left zoom in Fig.\ref{fig2}c, grey solid line). The free energy difference equals the reversible work of releasing the polypeptide chain from force $f$ to 0 (-$\Delta G_{\rm p}(0\to f)$) plus the reversible work of relaxing the handles and the bead in the optical trap from $f$ to zero (equal to -$\Delta G_{\rm h-b}(0\to f)$, from step 1$\rightarrow$2). Thermodynamic energy differences are path-independent so $\Delta G( 1\rightarrow 4) = \Delta G(1 \rightarrow 2) + \Delta G(2\rightarrow 3) + \Delta G(3\rightarrow 4)$. This gives,
\begin{equation}
    \Delta G_0 = \Delta G_{\rm d}(0\to f)+\Delta G(f) -\Delta G_{\rm p}(0\to f) 
\label{eqG}
\end{equation}
The same balance equation holds for $\Delta H_0$, $\Delta S_0$, $\Delta C_{p}$. Notice that $\Delta G_{\rm h-b}(0\to f)$ does not appear in Eq.\ref{eqG} as it cancels out when adding steps 1$\rightarrow$2 and 3$\rightarrow$4 (section S1 in Supp. Info. for details). 

The unfolding transition at constant force can be measured in instruments where the intensive variable, i.e., the force, is the natural control parameter (e.g., in magnetic tweezers). In contrast, in optical tweezers force cannot be controlled unless force feedback is applied \cite{MRico18}. As a consequence, the unfolding transition does not occur at fixed force $f$ but at fixed $\lambda$. Indeed, when pulling with optical tweezers the unfolding transition is observed as a force rip in the FDC (dark arrows in Fig.\ref{fig2}b) which occurs at fixed $\lambda$. Therefore, free energy differences in the force-ensemble, $\Delta G(f)$, are Legendre transforms of those measured in the $\lambda$-ensemble \cite{Amartinez18}.

A major contribution in Eq.\ref{eqG} is the elastic term $\Delta G_{\rm p}$ for the polypeptide chain, which is often modelled as a semiflexible polymer. The term $\Delta G_{\rm d}$ stands for the elastic energy of aligning a molecular-sized dipole along the force axis. As the dipole extension is much shorter than the contour length of the polypeptide chain, $\Delta G_{\rm p}(0\to f)\gg \Delta G_{\rm d}(0\to f)$ at all forces. The relative magnitude of $\Delta G_{\rm p}(0\to f)$ and $\Delta G(f)$ depends on the difference between $f$ and the coexistence force $f_c$, which is defined as the force at which the folded and unfolded barnase have equal free energies, i.e., $\Delta G(f_c) = 0$. Equation \ref{eqG} gives for $f=f_c$ ,
\begin{equation}
    \Delta G_0 = -\Delta G_{\rm p}(0\to f_c) + \Delta G_{\rm d}(0\to f_c) ~.
\label{eqG0}
\end{equation}
The stretching free energy of the different elastic elements at a given force $f$ can be obtained by using the well-known expression \cite{Amartinez18}
\begin{equation}
  \Delta G_{\rm i}(0\to f)=-\int_0^f x_{\rm i}(f')df'
  \label{eqDGI}
\end{equation}
where i$\equiv$p,d whereas the difference in the trap position $\Delta\lambda$ between the unfolded and native branches at a given force $f$ (right zoom in Fig.\ref{fig2}b) equals
\begin{equation}
   \Delta\lambda(f)= x_{\rm p}(f) - x_{\rm d}(f)\,\, .
   \label{eqAL}
\end{equation}
Combining the previous equations we obtain the relation,
\begin{equation}
    \Delta G_0(T)=\int_0^{f_c(T)}\Delta\lambda(f,T)df \ \rm ,
    \label{eqGint}
\end{equation}
showing that the knowledge of $f_c(T)$ and the measured $\Delta\lambda(f,T)$ permits to determine $\Delta G_0$ at a given temperature $T$. Equation \ref{eqGint} is the basic thermodynamic formula we will use to determine $\Delta S_0(T)$, $\Delta H_0(T)$ and $\Delta C_p$ in barnase folding.

\printbibliography
\end{document}


\maketitle

\section{Folding free energy derivation}
\subsection{Relation between $\Delta G(f)$ and $\Delta G_0$}

In CFS experiments the free energy difference at a given force, $\Delta G(f)$, is defined as
\begin{equation}\label{eq1.sec1}
    \Delta G(f) = G_{\rm U}(f) - G_{\rm N}(f) \ ,
\end{equation}
where $G_{\rm U}(f)$ and $G_{\rm N}(f)$ are the free energies of the unfolded (U) and native (N) states at force $f$, respectively. Taking the native protein at zero force as reference state,  $G_{\rm U}(f)$ and $G_{\rm N}(f)$ are calculated as in \cite{Amartinez18},
\begin{subequations}\label{eq2.sec1}
    \begin{align}
        & G_{\rm U}(f) = - \int_0^{f} \lambda_{\rm U}(f')df' + \Delta G_0 \\
        &  G_{\rm N}(f) = - \int_0^{f} \lambda_{\rm N}(f')df'
    \end{align}
\end{subequations}
with $\Delta G_0 (>0)$ the free energy of the random coil state relative to the native protein state. Here $\lambda_{\rm U}$ and $\lambda_{\rm N}$ are the extension of the system when the protein is in U and N at a given force $f$. Moreover, the extension of the system $\lambda$ is decomposed as:
\begin{subequations}\label{eq3.sec1}
    \begin{align}
        \lambda_{\rm U}(f) = x_{\rm b}(f) + x_{\rm h}(f) + x_{\rm p}(f) \\
        \lambda_{\rm N}(f) = x_{\rm b}(f) + x_{\rm h}(f) + x_{\rm d}(f)
    \end{align}
\end{subequations}
 where $x_{\rm b}(f)$ is the displacement of the bead with respect to the center of the optical trap; $x_{\rm h}(f)$ is the extension of the dsDNA handles; $x_{\rm p}(f)$ is the extension of the polypeptide chain; and, $x_{\rm d}(f)$ is the N-C termini distance when the protein is in N \cite{AAlemany_2016}.

Introducing Eqs.\ref{eq2.sec1}a,b to Eq.\ref{eq1.sec1} and taking into consideration Eqs.\ref{eq3.sec1}a,b we obtain,
\begin{subequations}\label{eq4.sec1}
    \begin{align}
         \Delta G(f) &= - \sum\limits_{\rm i=b,h,p}\int_0^{f} x_{\rm i}(f')df' + \Delta G_0 + \sum\limits_{\rm i=b,h,d}\int_0^{f} x_{\rm i}(f')df' = \\
         &= \Delta G_0 - \int_0^{f} x_{\rm p}(f')df' + \int_0^{f} x_{\rm d}(f')df'
    \end{align}
\end{subequations}
where the bead and handles contributions cancel out. The two integrals of $x_{\rm p}(f)$ and $x_{\rm d}(f)$ in Eq.\ref{eq4.sec1}b equal the free energies required to stretch the polypeptide chain and orient the native protein (dipole), respectively. Notice that Eq.\ref{eq4.sec1} is Eq.4 and Eq.11 in the main text. In a more compact form,
\begin{equation}\label{eq7.sec1}
        \Delta G(f) = \Delta G_0 - \int_0^{f} (x_{\rm p}(f')-x_{\rm d}(f'))df' = \Delta G_0 - \int_0^{f} \Delta\lambda(f')df'
\end{equation}
which for $f=f_c(T)$ gives Eq.1 and Eq.15 in the main text and $\Delta\lambda(f)$ is the difference between the extension of the native and unfolded states at force $f$, Eq.14 in the main text.

\subsection{Relation between $\Delta G_\lambda$ and $\Delta G_0$}

In pulling experiments the molecular construction is driven out of equilibrium from an initial position ($f_{\rm min},\lambda_{\rm min}$) where the molecule under study is always in its native state to a final position ($f_{\rm max},\lambda_{\rm max}$) where it is always unfolded. The energy difference between these two points, $\Delta G_\lambda$, is defined as
\begin{equation}\label{eq1.sec2}
    \Delta G_\lambda =  G_{\rm U}^{\lambda_{\rm max}} - G_{\rm N}^{\lambda_{\rm min}}
\end{equation}
where $G_{\rm U}^{\lambda_{\rm max}}$ and $G_{\rm N}^{\lambda_{\rm min}}$ are the free energies of the unfolded and native states relative to the native state at zero force (corresponding to $\lambda=0$) which is taken as reference state. Free energies in Eq.\ref{eq1.sec2} are the Legendre transforms of Eq.\ref{eq1.sec1},
\begin{subequations}\label{eq2.sec2}
    \begin{align}
      &  G_{\rm U}^{\lambda_{\rm max}} = G_{\rm U}^{f_{\rm max}} + \lambda_{\rm U}(f_{\rm max})f_{\rm max} \\
      &  G_{\rm N}^{\lambda_{\rm min}} = G_{\rm N}^{f_{\rm min}} + \lambda_{\rm N}(f_{\rm min})f_{\rm min} \ \rm .
    \end{align}
\end{subequations}
where $G_{\rm N/U}^{f_{\rm min}/f_{\rm max}}$ is obtained from Eq.\ref{eq2.sec1}a,b, 
\begin{subequations}\label{eq3.sec2}
    \begin{align}
      & G_{\rm U}^{f_{\rm max}} = - \int_0^{f_{\rm max}} \lambda_{\rm U}(f')df' + \Delta G_0 \\
      & G_{\rm N}^{f_{\rm min}} = - \int_0^{f_{\rm min}} \lambda_{\rm N}(f')df' \ \rm .
    \end{align}
\end{subequations}

Introducing Eqs.\ref{eq2.sec2}a,b and Eqs.\ref{eq3.sec2}a,b into Eq.\ref{eq1.sec2} we obtain 
\begin{equation}\label{eq4.sec2}
    \Delta G_\lambda = - \int_0^{f_{\rm max}} \lambda_{\rm U}(f')df' + \Delta G_0 +  \lambda_{\rm U}(f_{\rm max})f_{\rm max}  + \int_0^{f_{\rm min}} \lambda_{\rm N}(f')df' -  \lambda_{\rm N}(f_{\rm min})f_{\rm min} \rm ,
\end{equation}

which can be rewritten as  
\begin{equation}\label{eq5.sec2}
    \Delta G_\lambda = \Delta G_0 - \sum\limits_{i = \rm b,h,p} \left[ \int_0^{f_{\rm max}} x_{\rm i}(f')df' +   x_{\rm i}(f_{\rm max})f_{\rm max}  \right] + \sum\limits_{i = \rm b,h,d} \left[ \int_0^{f_{\rm min}} x_{\rm i}(f')df' -   x_{\rm i}(f_{\rm min})f_{\rm min}  \right]\ 
\end{equation}
considering the definitions of $\lambda_{\rm U}$ and $\lambda_{\rm N}$ in Eq.\ref{eq3.sec1}. Integrating by parts, we obtain the final expression for $\Delta G_\lambda$:
\begin{subequations}\label{eq7.sec2}
    \begin{align}
    \Delta G_\lambda & = \Delta G_0 + \sum\limits_{\rm i = b,h,p} \left[ \int_0^{x_{\rm i}(f_{\rm max})} f_{\rm i}(x')dx'   \right] - \sum\limits_{\rm i = b,h,d} \left[ \int_0^{x_{\rm i}(f_{\rm min})} f_{\rm i}(x')dx'   \right] = \\
    & = \Delta G_0 + \sum\limits_{\rm i = b,h} \left[ \int_{x_{\rm i}(f_{\rm min})}^{x_{\rm i}(f_{\rm max})} f_{\rm i}(x')dx'  \right] + \int_0^{x_{\rm p}(f_{\rm max})} f_{\rm p}(x')dx' - \int_0^{x_{\rm d}(f_{\rm min})} f_{\rm d}(x')dx'
    \end{align}
\end{subequations}
or, in abbreviated form,
\begin{subequations}\label{eq8.sec2}
    \begin{align}
 \Delta G_\lambda & = \Delta G_0 + W_{\lambda}^{\rm elas} \\
W_{\lambda}^{\rm elas} & = \Delta G_{\rm b}^{\lambda_{\rm min} \to \lambda_{\rm max}} + \Delta G_{\rm h}^{\lambda_{\rm min} \to \lambda_{\rm max}} + \Delta G_{\rm p}^{0 \to \lambda_{\rm max}} - \Delta G_{\rm d}^{0 \to \lambda_{\rm min}} 
    \end{align}
\end{subequations}
which is Eq.5 in the main text. The different terms in Eq.\ref{eq8.sec2}b are defined as
\begin{subequations}\label{eq9.sec2}
    \begin{align}
\Delta G_{\rm b}^{\lambda_{\rm min} \to \lambda_{\rm max}} =  \int_{x_{\rm b}(f_{\rm min})}^{x_{\rm b}(f_{\rm max})} f_{\rm b}(x')dx' \\
\Delta G_{\rm h}^{\lambda_{\rm min} \to \lambda_{\rm max}} =  \int_{x_{\rm h}(f_{\rm min})}^{x_{\rm h}(f_{\rm max})} f_{\rm h}(x')dx' \\
\Delta G_{\rm p}^{0 \to \lambda_{\rm max}} =  \int_0^{x_{\rm p}(f_{\rm max})} f_{\rm p}(x')dx' \\
\Delta G_{\rm d}^{0 \to \lambda_{\rm min}} =  \int_0^{x_{\rm b}(f_{\rm min})} f_{\rm d}(x')dx'
    \end{align}
\end{subequations}
where $f_{\rm i}(x)$ is the inverse function of the elastic response, $x_{\rm i}(f)$, of the different elements (b, h, p, d).

\newpage
\begin{figure}[h]
    \centering
    \includegraphics[width = 0.7\textwidth]{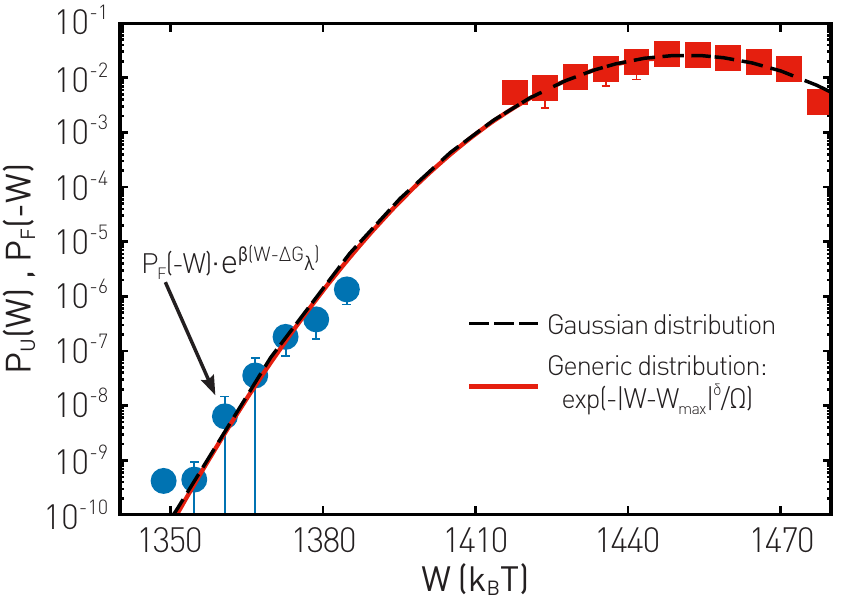} 
    \caption{{\bf Matching approach.} Experimental unfolding work distribution (red squares) and folding work distribution multiplied by $\exp{\Bigl( \frac{W-\Delta G_\lambda}{k_BT}\Bigr)}$ (blue circles) at 25ºC for one illustrative molecule together with the function $exp[(-|W-W_{max} |^\delta/ \Omega)]$ (with $\delta \sim$1.9) (red line) and the Gaussian distribution (black dashed line) to demonstrate that both distributions hold to the same $\Delta G_\lambda$. The fact that the generic and the Gaussian distribution are nearly the same demonstrates that $\delta \sim$ 2 is an excellent approximation to the tails of $P_U (W)$ and $P_F (-W)$ around $W=\Delta G_\lambda$.} 
    \label{figS1}
\end{figure}

\newpage

\begin{figure}[h]
    \centering
    \includegraphics[width = 0.8\textwidth]{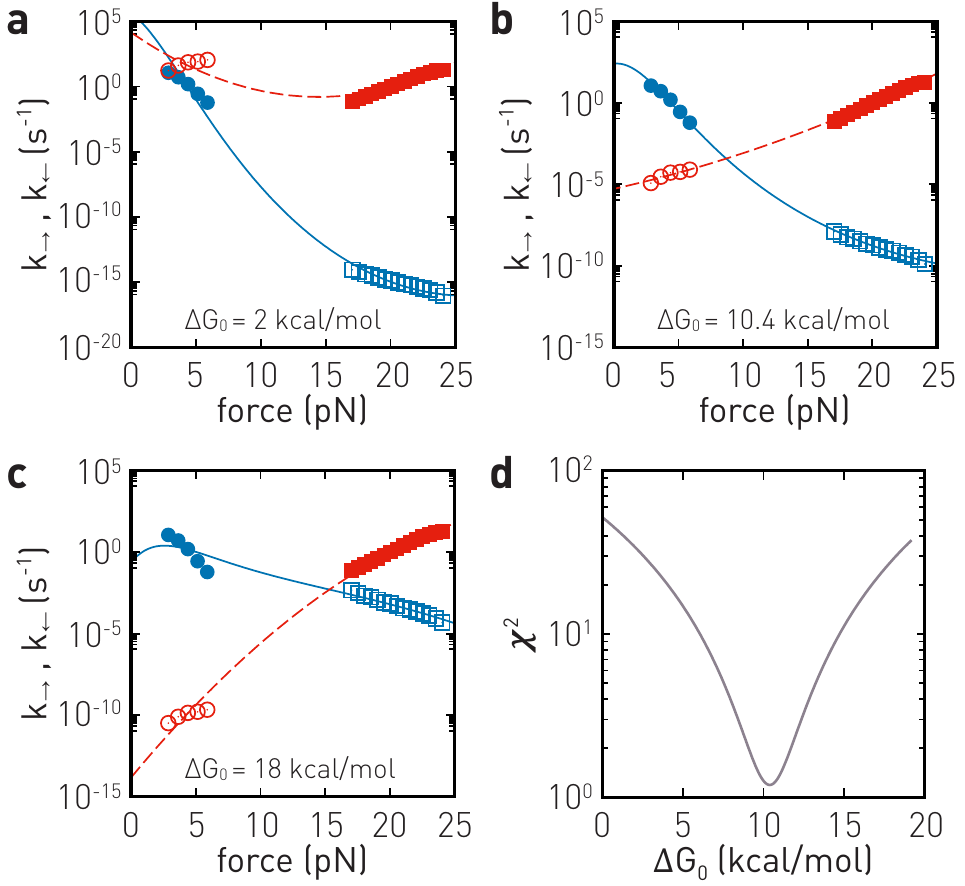}
    \caption{Unfolding (red symbols) and folding (blue symbols) kinetic rates at 298K. \textbf{a)}, \textbf{b)}, and \textbf{c)} Measured (solid symbols) and extrapolated (empty symbols) using the detailed balance condition kinetic rates. \textbf{d)} $\chi^2$  between the predicted force-dependent kinetic rates (lines) and the experimental and reconstructed rates (symbols) as a function of the $\Delta G_0$ value. The value of $\Delta G_0$ corresponding to the minimum $\chi^2$ gives our best estimate for the folding free energy ($\Delta G_0 = 10.4$kcal/mol, panel b). This value agrees better with that derived from the work-FT ($\Delta G_0  =10.3$kcal/mol, from Figure \ref{figS1} after subtracting elastic contributions).}
    \label{figS2}
\end{figure}

To determine $\Delta G_0$ and the force-dependent kinetic rates, we proceed as illustrated in Figure \ref{figS2} for the room temperature case, $T=298$K. First, we use the detailed balance condition, $k_\leftarrow(f)/k_\to(f) = \exp(\Delta G(f)/k_BT)$, to reconstruct the unfolding (folding) kinetic rates in the region of forces where folding (unfolding) transition events are observed in the pulling experiments. We do this for a given value of $\Delta G_0$ and the elastic properties derived at that temperature. It is important to note that such a reconstruction (which depends on the value of $\Delta G_0$) is done in the force ensemble, therefore, the elastic contributions of bead and handles do not contribute to the overall free energy change $\Delta G(f)$, and only the energy to stretch the polypeptide chain and orient the folded protein contribute to $\Delta G(f)$. The explanation of this calculation follows the steps described in section “Folding free energy at zero force” in Materials and Methods.

The reconstructed kinetic rates at $T=298$K are shown in panels a, b, and c of Figure \ref{figS2} for three selected values of $\Delta G_0$ ($2$, $10.4$, and $18$ kcal/mol, respectively). As we can see, too low (panel a) or too high (panel c) values of $\Delta G_0$ do not permit to analytically continue the force-dependent kinetic rates from the measured force region to the reconstructed force region. For example, Fig. \ref{figS2}a shows that it is not possible to analytically continue the measured unfolding kinetic rates at high forces (filled red squares) to the reconstructed rates at low forces (empty red circles) without assuming a strongly non-monotonic force dependence. Concomitantly, it is not possible to continue the measured folding kinetic rates at low forces (filled blue circles) to the reconstructed rates at high forces (empty blue squares). To estimate the quality of the analytical continuation of the rates between the two force regimes, we have fitted all unfolding kinetic rates (measured and reconstructed) to a single quadratic function in the log-normal plot (dashed red lines). From the dashed red lines and using detailed balance, we inferred the folding kinetic rates (solid blue lines). Notice that the latter are not necessarily quadratic functions. Next, we have calculated the overall mean-square deviation ($\chi^2$) between the predicted force-dependent kinetic rates (dashed lines) and the experimental plus reconstructed ones (symbols). This has been done by varying the value of $\Delta G_0$ at regular steps of $0.2$kcal/mol (Figure \ref{figS2}d). The value of $\Delta G_0$ with minimum $\chi^2$ gives our best estimate for the folding free energy ($\Delta G_0 = 10.4 \pm 0.2$ kcal/mol, panel b).

\newpage

\begin{figure}[h]
    \centering
    \includegraphics[width = 0.9\textwidth]{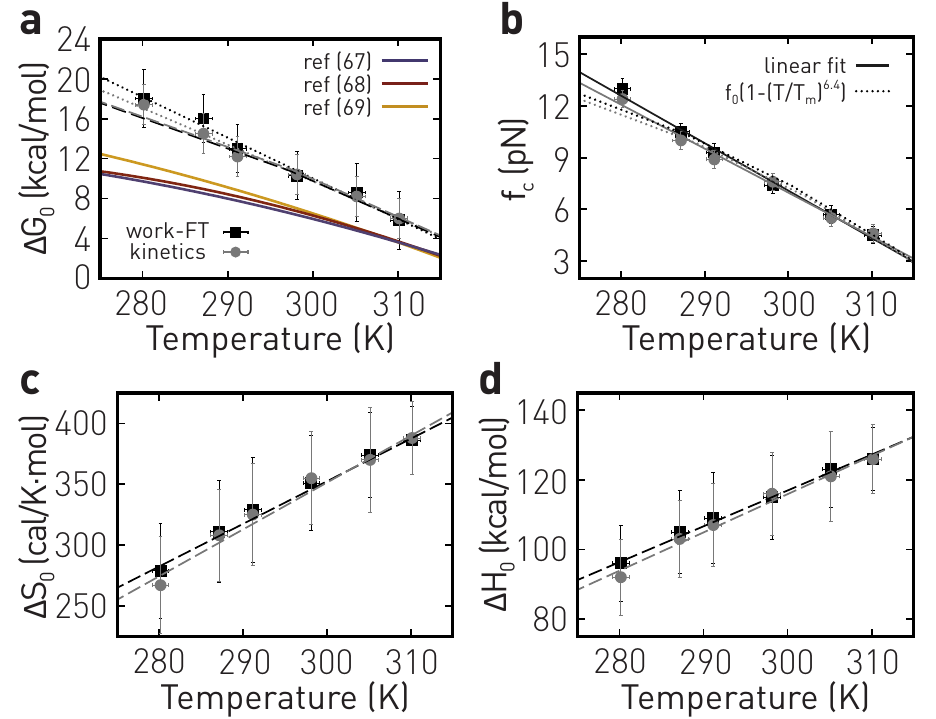}
    \caption{{\bf Comparison between Work-FT and kinetics. } \textbf{(a)} Folding free energy derived from the Work-FT (black squares); and folding free energy difference derived from the kinetics using the detailed balance condition (gray circles). Dotted lines correspond to linear fits, and dashed lines correspond to the protein stability curve. The solid colored (gray, brown, yellow) lines are derived from the references (67), (68), and (69). \textbf{(b)} Coexistence force determined using the $\Delta G_0$ value derived from the work-FT (black squares) and Eq.1 of the main text; and  coexistence force measured as the crossing point of the kinetic rates (gray circles). Solid lines are linear fits, while dotted lines are fits to $f_0 (1-(T/T_m )^\alpha)$ imposing $\alpha=6.4$ and the melting temperature. \textbf{(c)} Entropy of folding derived from the coexistence force shown in \textbf{b} and Eq.2 of the main text. \textbf{(d)} Enthalpy of folding obtained from the measured values of $\Delta G_0$ and $\Delta S_0$.}
    \label{figS3}
\end{figure}

In Figure \ref{figS3} we compare the results of $\Delta G_0(T)$, $f_c(T)$, $\Delta S_0(T)$ and, $\Delta H_0(T)$ obtained using the work-FT (section D in the main text) or the kinetic rates (kinetics, section E in the main text). In panel b, the measured $f_c(T)$ values have been fitted to the formula $f_0 (1-(T/T_m )^\alpha)$ studied to investigate the critical forces in \cite{klimov1999, hyeon2005, imparato2009} (dotted lines). The parameter $\alpha$ is fixed to $\alpha = 6.4$ as reported in \cite{hyeon2005} and $T_m$ is fixed to the melting temperature measured from panel a.

%

\newpage

\begin{figure}[h]
    \centering
    \includegraphics[width = 0.8 \textwidth]{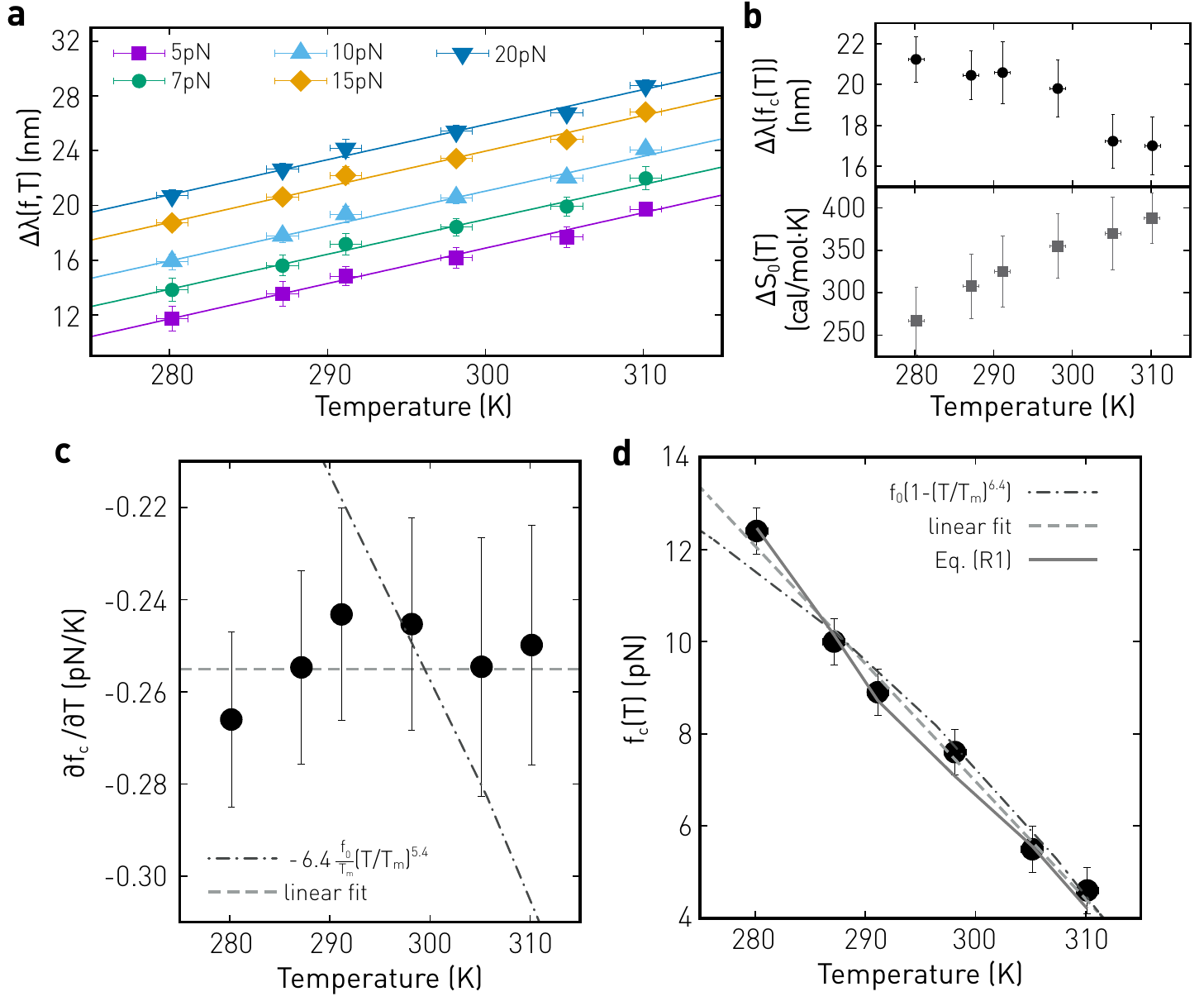}
    \caption{{\bf Coexistence force line. } \textbf{(a)} Trap position difference, $\Delta \lambda(f,T)$, measured at different forces (Fig. 3 in main text): 5pN (purple), 7pN (green), 10pN (cyan), 15pN (yellow), and 20pN (blue) at different temperatures. Solid lines are linear fits to the experimental data.  \textbf{(b)} Top: Measured $\Delta \lambda(f,T)$ at the coexistence force $f_c (T)$ (mean value obtained from the fluctuation theorem and the kinetics, Fig. 4d and 5d-top, in main text): $\Delta \lambda(f_c(T))\equiv \Delta \lambda(f_c (T),T)$. Bottom: Measured average entropy difference at zero force, $\Delta S_0(T)$ (mean value obtained from the fluctuation theorem and the kinetics, Figs. 4e and 5d-bottom in main text). c) Points are the measured $\partial f_c/ \partial T$ from \eqref{inv_Eq2} using the values shown in panels \textit{a}, and \textit{b}. Dashed-dotted line corresponds to the derivative of the function $f_c(T)=f_0 (1-(T/T_m)^{6.4})$ fitted in panel \textit{d} (see also Fig. \ref{figS3}b). The horizontal dashed line shows the slope of the linear fit done in panel \textit{d}. \textbf{(d)} Coexistence force of barnase as a function of temperature, $f_c(T)$. Black dashed-dotted line is a fit to the function $f_c(T)=f_0 (1-(T/T_m)^{6.4})$, dashed line corresponds to a linear fit, solid line is the reconstructed temperature behavior using the measured $\partial f_c/\partial T$ measured using \eqref{inv_Eq2} (black circles in \textit{c}). }
    \label{figS4}
\end{figure}
Regarding the linear fit to $f_c(T)$, we just chose the simplest behavior that fits the data. If the transition is first order and the force and temperature do not span a too wide range, then one should naively expect a linear $f-T$ coexistence behavior (as observed in the data). This is a consequence of the Clausius-Clapeyron equation (Eq. (2) in the main text) which predicts that the slope of the $f-T$ line equals the total latent heat of unfolding the protein and stretching the polypeptide chain, divided by the change in molecular extension $\Delta\lambda$ across the transition (in main text, Fig. 4e and 5d-bottom for $\Delta S_0$, and Fig. 3 for $\Delta\lambda$),
\begin{equation}\label{inv_Eq2}
    \frac{\partial f_c}{\partial T}=-\frac{\Delta S_0 (T)+\int_0^{f_c (T)}\partial \Delta \lambda(f',T)/\partial T  df' }{\Delta \lambda(f_c (T))} 
\end{equation}
Figure \ref{figS4} summarizes the results obtained by testing \eqref{inv_Eq2}. From Fig. 3 of the main text, we have calculated $\Delta \lambda(f,T)$ versus $T$ at different forces (Fig. \ref{figS4}a). Using the values of $f_c(T)$ (mean value obtained from the fluctuation theorem and the kinetics, Figs. 4d and 5d-top in main text), we have calculated $\Delta \lambda(f_c (T))\equiv \Delta\lambda(f_c (T),T)$ (Fig. \ref{figS4}b-top). In Figure \ref{figS4}b-bottom, we show the measured average entropy difference at zero force, $\Delta S_0 (T)$ (mean value obtained from the fluctuation theorem and the kinetics, Fig. 4e and 5d-bottom, main text). By inserting $\Delta \lambda(f,T)$, $\Delta S_0 (T)$, and $\Delta\lambda(f_c (T))$ in \eqref{inv_Eq2}, we have calculated the derivative $\partial f_c (T)/\partial T$ in the range 7-37ºC, finding the results shown in Fig. \ref{figS4}c. As we can see, $\partial f_c (T)/\partial T$ is nearly constant in agreement with the linear prediction, $f_c (T)=a(T-T_m)$. The studied linear function fits better the data than the function $f_c (T)=f_0 (1-(T/T_m )^{6.4})$ (Figs. \ref{figS4}c and \ref{figS4}d). 
 
\newpage

\begin{figure}[h]
    \centering
    \includegraphics[width = 0.9\textwidth]{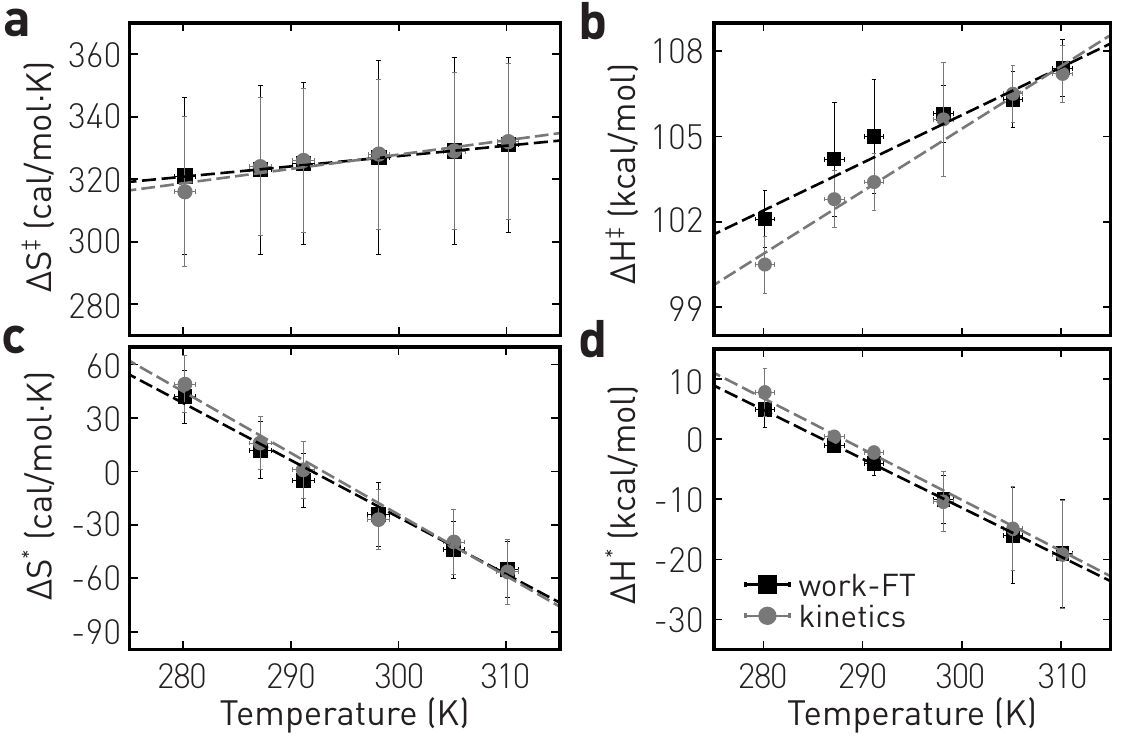}
    \caption{{\bf Entropy and Enthalpy of the TS.} \textbf{(a)} Entropy and \textbf{(b)} Enthalpy differences between N and TS using the values of $\Delta S_0$ and $\Delta H_0$ derived from work-FT (black squares) or the kinetics (gray circles). \textbf{(c)} Entropy and \textbf{(d)} Enthalpy differences between U and TS using the values of $\Delta S_0$ and $\Delta H_0$ derived from work-FT (black squares) or the kinetics (gray circles). }
    \label{figS5}
\end{figure}

\newpage

\begin{figure}[h]
    \centering
    \includegraphics[width = 0.9\textwidth]{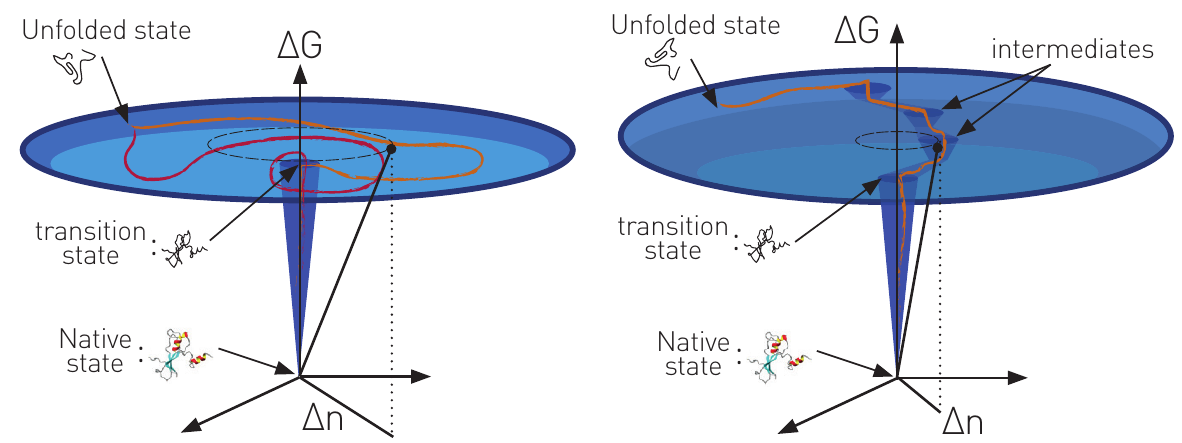}
    \caption{{\bf Funneled free energy landscapes.} These can be represented in cylindrical-like coordinates with the vertical axis denoting the free energy relative to N and the radial distance to the vertical axis indicating the change in number of degrees of freedom $\Delta n$ relative to N. The major change in $\Delta G$ occurs between N and TS along the vertical direction, while the major change in $\Delta n$ occurs between TS and U along the radial basin. Left: Golf-course free energy landscape compatible with the ELH. The orange and red lines denote two folding pathways from U to N passing through TS. Right: Golf-course free energy landscape compatible with the FH. Notice the well-defined folding pathway connecting intermediates states (orange line). In both free energy landscapes are remarked the unfolded, transition, and native states.  }
    \label{figS6}
\end{figure}

\newpage

\begin{table}[t]
\begin{tabular}{c||c|c|c|c|c|c|c}
                         & pH  & \begin{tabular}[c]{@{}c@{}}$T_m$ \\ (ºC)\end{tabular}  & \begin{tabular}[c]{@{}c@{}}ionic \\ strength\end{tabular}   & \begin{tabular}[c]{@{}c@{}}$\Delta G_0^{\ 298K}$ \\ (kcal/mol)\end{tabular} & \begin{tabular}[c]{@{}c@{}}$\Delta H_0^m$ \\ (kcal/mol)\end{tabular} & \begin{tabular}[c]{@{}c@{}}$\Delta S_0^m$ \\ (cal/K$\cdot$mol) \end{tabular} & \begin{tabular}[c]{@{}c@{}}$\Delta C_p$ \\ (cal/K$\cdot$mol)\end{tabular} \\ \hline 
\textbf{This study} & \textbf{7.0} & \textbf{50 $\pm$ 2} & \textbf{20 mM}  & \textbf{10 $\pm$ 2} & \textbf{140 $\pm$ 2} & \textbf{434 $\pm$ 10} & \textbf{1050 $\pm$ 50}  \\ \hline 
(a) & 5.0-8.0 & 52 &  & 11.8 & 140 & 430 &  \\ \hline \hline
(b) & 6.3 & 53.9 &  & 9.4 & 125 & 385 &  \\
(b) & 6.3 & 53.9 &  & 8.9 & 125 & 385 &  \\
(b) & 6.3 & 53.9 &  & 9.5 & 125 & 385 &  \\ \hline \hline
(c) & 6.3 &  & 50 mM & 9.9 &  &  &  \\ \hline
(d) & 6.3 &  & 10 mM & 10.0 $\pm$ 0.6 &  &  &  \\
(d) & 6.3 &  & 50 mM & 10.2 $\pm$ 0.6 &  &  &  \\ \hline \hline
(e) & 6.3 &  & 7 mM & 8.5 &  &  &  \\
(e) & 6.3 &  & 10 mM & 9.0 &  &  &  \\
(e) & 6.3 &  & 50 mM & 10.3 &  &  &  \\ \hline \hline
(f) & 7.0 &  & 30 mM & 8.9 &  &  &  \\
(f) & 7.0 &  & 30 mM & 8.5 &  &  &  \\ \hline
(g) & 5.9 &  & 30 mM & 8.1 &  &  &  \\
(g) & 8.9 &  & 30 mM & 8.0 &  &  &  \\ \hline \hline
(h) & 4.5 & 52.4 &  &  & 120 $\pm$ 7 & 367 & 1360 $\pm$ 140 \\
(h) & 4.9 & 54.6 &  &  & 120 $\pm$ 7 & 367 & 1360 $\pm$ 140 \\
(h) & 5.5 & 53.8 &  &  & 120 $\pm$ 7 & 367 & 1360 $\pm$ 140 \\ \hline
(i) & 6.3 & 54.0 & 50 mM & 9.3 $\pm$ 0.8 &  &  &  \\
(i) & 6.3 & 54.0 & 50 mM & 9.1 $\pm$ 0.5 &  &  &  \\
(i) & 6.3 & 54.0 & 50 mM & 9.1 $\pm$ 0.3 &  &  &  \\
(i) & 6.3 & 54.0 & 50 mM & 9.1 $\pm$ 0.4 &  &  &  \\
(i) & 6.3 & 54.0 & 50 mM & 9.3 $\pm$ 0.5 &  &  &  \\
(i) & 6.3 & 54.0 & 50 mM & 8.8 $\pm$ 0.15 &  &  &  \\ 
(j) & 5.5-5.8 & 50.0 &  &  & 107 $\pm$ 6 & 330 $\pm$ 18 & 860 \\ \hline
(k) & 6.2 &  & 10 mM &  & 135 & 415 & 1360 $\pm$ 140 \\ \hline
(l) & 4.5-5.0 & 53.0 &  & 9.3 & 130 $\pm$ 1 & 400 $\pm$ 4 & 1480 $\pm$ 190 \\ \hline
(m) & 6.3 &  & & 8.9 $\pm$ 0.1 &  &  & 1600 $\pm$ 170  \\
(m) & 6.3 &  &  & 10.5 $\pm$ 0.1 &  &  & 1600 $\pm$ 170  \\ \hline
(n) & 6.0 &  & 50 mM & 10.0 & 140 $\pm 20$ & 430 $\pm$ 61 & 1880 $\pm$ 700 \\ \hline \hline

\end{tabular}
\caption{Thermodynamic data collection. References: \textbf{(a)} R. W. Hartley, (1969) Biochemistry 8(7), 2929–2932. \textbf{(b)} J. T. Kellis Jr, K. Nyberg, and A. R. Fersht, (1989) Biochemistry 28(11),4914–4922. \textbf{(c)} A. Horovitz, L. Serrano, B. Avron, M. Bycroft, and A. R. Fersht, (1990) Journal of molecular biology 216(4), 1031–1044. \textbf{(d)} L. Serrano, A. Horovitz, B. Avron, M. Bycroft, and A. R. Fersht, (1990) Biochemistry 29(40), 9343–9352. \textbf{(e)} M. Bycroft, A. R. Fersht, et al., (1991) Journal of molecular biology 220(3), 779–788.  \textbf{(f)} C. N. Pace, D. V. Laurents, and R. E. Erickson, (1992) Biochemistry 31(10), 2728–2734. \textbf{(g)} J. Sancho, L. Serrano, and A. R. Fersht, (1992) Biochemistry 31(8), 2253–2258. \textbf{(h)} ref (65) in main text \textbf{(i)} S. Vuilleumier, J. Sancho, R. Loewenthal, and A. R. Fersht, (1993) Biochemistry, 32(39), 10303–10313., \textbf{(j)} ref (64) in main text \textbf{(k)} A. Makarov, I. Protasevich, V. Lobachov, M. Kirpichnikov, G. Yakovlev, R. Gilli, C. Briand, and R. Hartley, (1994) FEBS letters 354(3), 251–254. \textbf{(l)} J. C. Martinez, M. E. Harrous, V. V. Filimonov, P. L. Mateo, and A. R. Fersht, (1994) Biochemistry 33(13), 3919–3926. \textbf{(m)} A. Matouschek, J. M. Matthews, C. M. Johnson, and A. R. Fersht, (1994) Protein Engineering, Design and Selection 7(9), 1089–1095. \textbf{(n)} M. Oliveberg, S. Vuilleumier, and A. R. Fersht, (1994) Biochemistry 33(29), 8826–8832.}
\end{table}

\begin{table}[t]
\begin{tabular}{c||c|c|c|c|c|c|c}
                         & pH  & \begin{tabular}[c]{@{}c@{}}$T_m$ \\ (ºC)\end{tabular}  & \begin{tabular}[c]{@{}c@{}}ionic \\ strength\end{tabular}   & \begin{tabular}[c]{@{}c@{}}$\Delta G_0^{\ 298K}$ \\ (kcal/mol)\end{tabular} & \begin{tabular}[c]{@{}c@{}}$\Delta H_0^m$ \\ (kcal/mol)\end{tabular} & \begin{tabular}[c]{@{}c@{}}$\Delta S_0^m$ \\ (cal/K$\cdot$mol) \end{tabular} & \begin{tabular}[c]{@{}c@{}}$\Delta C_p$ \\ (cal/K$\cdot$mol)\end{tabular} \\ \hline 
\textbf{This study} & \textbf{7.0} & \textbf{50 $\pm$ 2} & \textbf{20 mM}  & \textbf{10 $\pm$ 2} & \textbf{140 $\pm$ 2} & \textbf{434 $\pm$ 10} & \textbf{1050 $\pm$ 50}  \\ \hline 
(o) & 6.3 &  &  & 10.5 $\pm$ 0.1 &  &  &  \\ \hline
(p) &  &  & 54.0 &  & 112 & 330 & 1410 \\ \hline
(q) & 7.0 & 54.4 $\pm$ 0.5 &  &  & 131 $\pm$ 5 & 403 $\pm$ 15 &  \\ \hline
(r) & 6.3 & 53.6 &  & 8.9 $\pm$ 0.3 &  &  &  \\ \hline
(s) & 6.3 & 54.8 & 50 mM & 10.2 $\pm$ 0.2 &  &  &  \\ \hline \hline
(t) & 6.3 & 49.6 & 50 mM &  & 145 & 445 &  \\
(u) & 6.3 & 53.9 & 50 mM &  &  &  & 1410 \\ \hline
(v) & 6.3 & 51.32 $\pm$ 0.003 & 50 mM & 10.3 & 141 & 434 &  \\ \hline \hline
(w) & 6.3 & 53.0 & 50 mM & 10.5  & 141  & 434 & 1690  \\ \hline \hline
(x) & 6.2 & 54.0 & 60 mM & 11.5 $\pm$ 1 & 129 $\pm$ 1 & 394 $\pm$ 3 &   \\
\hline \hline

\end{tabular}
\caption{Thermodynamic data collection. References: \textbf{(o)} C. M. Johnson and A. R. Fersht, (1995) Biochemistry 34(20), 6795–6804. \textbf{(p)} G. I. Makhatadze and P. L. Privalov, (1995) Advances in protein chemistry 47, 307–425. \textbf{(q)} J. C. Martinez, V. V. Filimonov, P. L. Mateo, G. Schreiber, and A. R. Fersht, (1995) Biochemistry 34(15), 5224–5233. \textbf{(r)} J. M. Matthews and A. R. Fersht, (1995) Biochemistry 34(20),6805–6814. \textbf{(s)} M. Oliveberg, V. L. Arcus, and A. R. Fersht, (1995) Biochemistry 34(29), 9424–9433. \textbf{(t)} W. S. Kwon, N. A. Da Silva, and J. T. Kellis Jr, (1996) Protein Engineering, Design and Selection 9(12), 1197–1202. \textbf{(u)} S. Gorinstein, M. Zemser, and O. Paredes-López, (1996) Journal of agricultural and food chemistry 44(1), 100–105. \textbf{(v)} R. Zahn, S. Perrett, and A. R. Fersht, (1996) Journal of molecular biology 261(1), 43–61. \textbf{(w)} P. A. Dalby, M. Oliveberg, and A. R. Fersht, (1998) Journal of molecular biology 276(3), 625–646. \textbf{(x)} ref (43)  in main text.}
\end{table}

\FloatBarrier
\printbibliography